\documentclass{aa}  
\usepackage{natbib}
\usepackage{xcolor}
\usepackage{graphicx}
\usepackage{txfonts}
\usepackage{hyperref}
\usepackage{float}
\usepackage{tablefootnote}
\usepackage{lipsum}
\usepackage{caption}
\usepackage{afterpage}
\usepackage{subcaption}
\usepackage{multirow}
\usepackage{lipsum}
\usepackage{arydshln}

\begin{document}

   \title{Astrometry-only detection of microlensing events with \textit{Gaia}}

   \author{T. Jankovi\v{c} \inst{1,2}
            \and
          A. Gomboc \inst{2}
          \and
          Ł. Wyrzykowski \inst{3}
          \and
          U. Kostić \inst{4}
          \and
          M. Karlica \inst{2,5}
          \and
          M. Larma \inst{2,6}  
        \and
        T. Petrushevska \inst{2}
          \and
          M. Bronikowski \inst{2}
          \and
          M. Jab{\l}o{\'n}ska \inst{3,7}
          \and
          Z. Kaczmarek \inst{8}
          }

   \institute{Institute of Physics of the Czech Academy of Sciences, Na Slovance 1999/2,182 21 Praha 8, Prague, Czech Republic\\
              \email{jankovic@fzu.cz}
        \and
        Center for Astrophysics and Cosmology, University of Nova Gorica, Vipavska 11c, 5270 Ajdov\v{s}\v{c}ina, Slovenia
         \and
            Astronomical Observatory, University of Warsaw, Al. Ujazdowskie 4, 00-478 Warszawa, Poland
      \and
           Aalta Lab d.o.o., Soška cesta 17, 5250 Solkan, Slovenia
       \and
            Astronomical Observatory, Volgina 7, 11000 Belgrade, Serbia
        \and
            Argelander-Institut für Astronomie, Universität Bonn, Auf dem Hügel 71, 53121 Bonn, Germany
        \and
            Research School of Astronomy and Astrophysics, Australian National University, Mount Stromlo Observatory, Cotter Road Weston Creek, ACT 2611, Australia
            \and
            Zentrum f{\"u}r Astronomie der Universit{\"a}t Heidelberg, Astronomisches Rechen-Institut, M{\"o}nchhofstr. 12-14, 69120 Heidelberg, Germany
             }

   \date{Received XX; accepted XX}

  \abstract
   {Astrometric microlensing events occur when a massive object passes between a distant source and the observer, causing a shift of the light centroid. The precise astrometric measurements of the \textit{Gaia} mission provide an unprecedented opportunity to detect and analyse these events, revealing properties of lensing objects such as their mass and distance.}
   {We develop and test the \textit{Gaia} Astrometric Microlensing Events (GAME) Filter, a software tool to identify astrometric microlensing events and derive lensing object properties.}
   {We generated mock \textit{Gaia} observations for different magnitudes, number of \textit{Gaia} visits, and events extending beyond \textit{Gaia}'s observational run. We applied GAME Filter to these datasets and validated its performance. We also assessed the rate of false positives where binary astrometric systems are misidentified as microlensing events.}
   {GAME Filter successfully recovers microlensing parameters for strong events. Parameters are more difficult to recover for short events and those extending beyond \textit{Gaia}’s run, where only a fraction of the events is observed. The astrometric effect breaks the degeneracy in the microlensing parallax present in photometric microlensing. For fainter sources, the observed signal weakens, reducing recovered events and increasing parameter errors. However, even for Gaia G-band magnitude 19, the parameters can be recovered for Einstein radii $\gtrsim 2\,${mas}. Observing regions with varying numbers of \textit{Gaia} visits has minimal impact on filter accuracy when the number of visits $\gtrsim 90$. Additionally, even if the peak of a microlensing event lies outside \textit{Gaia}’s run, microlensing parameters can still be recovered. GAME filter characterises lenses with astrometry-only data for lens masses ranging from approximately 1 to 20$\,\mathrm{M_\odot}$ and distances up to $\approx 6\,$kpc.}
   {GAME Filter reveals the astrometric potential in microlensing events. Our results indicate that astrometric data can enhance photometric detections and refine constraints on microlensing parameters.}

   \keywords{Astrometry - Gravitational lensing: micro - Methods: numerical}

\titlerunning{Astrometry-only detection of microlensing events with \textit{Gaia}}
   \maketitle

\section{Introduction}

Gravitational lensing, a phenomenon predicted by Einstein's General Theory of Relativity, occurs when a massive object warps spacetime and bends the light from a background source. This effect can create multiple images, amplify, and distort the source, making it a valuable tool for examining various astrophysical phenomena such as dark matter, distant galaxies, exoplanets, and stellar remnants \citep{Einstein1936, Refsdal1964}. When a stellar mass lensing object moves relative to the source, the deflection is usually under a few milliarcseconds and can last from days to years, resulting in a microlensing event. There are two primary types of microlensing: astrometric, which detects slight positional changes of the background source, and photometric, which observes variations in the source's brightness over time \citep{Paczynski1986, Hog1995, Miyamoto1995, Walker1995}.

Photometric microlensing has been extensively studied by decades-long surveys such as the Optical Gravitational Lensing Experiment (OGLE; \citealt{Udalski2003}) or Microlensing Observations in Astrophysics (MOA; \citealt{Bond2001}). However, only recently have technological advances and high-resolution observatories enabled the observation of astrometric microlensing \citep{Dominik_sahu_2000ApJ...534..213D, Belokurov2002} and allowed the detection and characterisation of objects that are difficult to observe through photometric methods alone, such as low-mass stars, white dwarfs, neutron stars, and black holes \citep{Sahu2017,Zurlo2018,mcgill2018, Dong2019, Sahu2022, Jablonska2022, Fardeen_2024ApJ...965..138F}. 

The \textit{Gaia} mission \citep{Gaia_mission}, launched by the European Space Agency, has revolutionised astrometry by providing high-precision measurements of the positions, motions, and parallaxes of more than a billion stars \citep{GaiaDR2, GaiaDR3, GaiaDR3b}. \textit{Gaia}'s precise astrometric data are particularly suited for identifying microlensing events. \citet{Belokurov2002} estimated that approximately 25,000 stars will exhibit noticeable centroid shifts during the mission. \textit{Gaia} results are published in Data Releases (DRs), which can be used to predict, find, or confirm astrometric microlensing events. Specifically, \citet{Kluter2018, Kluter2022} and \citet{Su2023} used \textit{Gaia} {data} to predict between 4,500 and 5,672 astrometric microlensing events from 2010 to 2070.

Despite the high number of expected astrometric microlensing events detected by \textit{Gaia}, confirmed observations remain limited to a single candidate. This candidate was identified by \citet{Jablonska2022} among the 363 photometric microlensing events in \textit{Gaia} DR3 \citep{Wyrzykowski2023}. With the upcoming  \textit{Gaia} {Data Release 4} in 2026, which will provide the complete astrometric time-series for all stars, there is a growing requirement for software tools to filter and detect astrometric microlensing events. By not relying on photometric data, such software would allow us to detect and study a larger sample of microlensing events, thereby enhancing our understanding of lensing objects. {\citet{Chen_2023} developed a data analysis pipeline for detecting astrometric microlensing events caused by black holes, with an emphasis on obtaining dark matter constraints. Their approach focuses on the detection of black holes using astrometry-only data of Gaia DR4, while in our work, we generalise the approach to other types of remnants. 
}

In this paper, we present the development and application of the \textit{Gaia} Astrometric Microlensing Events (GAME) Filter. This software tool is designed to identify microlensing events within \textit{Gaia}'s dataset and derive the properties of the lensing objects. We tested and validated the performance of GAME Filter using mock \textit{Gaia} observations across a range of magnitudes, numbers of observations, and durations of events. Additionally, we assessed the rate of false-positive detections involving binary systems misclassified as microlensing events. As it will be shown in this paper,  GAME Filter successfully recovers microlensing parameters. This indicates that GAME Filter holds great potential for providing insights into microlensing events and their associated lensing objects. {We note that, at present, Gaia data has been publicly released only up to DR3, which provides astrometric solutions including positions, proper motions, and parallaxes, but does not include the underlying astrometric time-series data necessary for direct microlensing detection. However, DR4, scheduled for release in 2026, will provide these detailed astrometric measurements, enabling the detection of astrometric microlensing events. Consequently, in this study, we validate GAME Filter using mock Gaia DR4 datasets.}

The paper is structured as follows. In Section \ref{sec:method}, we describe the basics of astrometric microlensing, the process of generating mock \textit{Gaia} observations, and the main properties of GAME filter. In Section \ref{sec:results}, we present the results, focusing on the performance of GAME Filter under various conditions.  Section \ref{sec:discussion} provides a discussion, and Section \ref{sec:conclusion} our main findings.

\section{Methodology}
\label{sec:method}

In this section, we present the main astrometric microlensing relations. We describe how we generated mock \textit{Gaia} observations with \texttt{astromet}\footnote{\href{https://github.com/zpenoyre/astromet.py}{https://github.com/zpenoyre/astromet.py}} and the main functionalities of GAME filter adapted for \texttt{jaxtromet}\footnote{\href{https://github.com/maja-jablonska/jaxtromet.py}{https://github.com/maja-jablonska/jaxtromet.py}}. \texttt{astromet} and \texttt{jaxtromet} are Python packages developed to simulate astrometric tracks of single sources, microlensing events, and binary systems (e.g., \citealt{Penoyre_2022MNRAS.513.2437P}). \texttt{astromet} is optimised for parallel CPU processing, while \texttt{jaxtromet} leverages the Python library JAX\footnote{\href{https://github.com/google/jax}{https://github.com/google/jax}} to perform calculations on GPUs, enhancing computational efficiency. {Simulating 50,000 mock \textit{Gaia} observations requires approximately 8 hours, while filtering and identifying microlensing events takes an additional 13 hours on a CPU with 24 cores}.

\subsection{Astrometric microlensing}

When a massive object passes near the line of sight to a source, a microlensing event occurs, creating two images of the source: a bright image close to the unlensed source position and a fainter image close to the lens. When the source is perfectly aligned with the lens, both images merge into a so-called Einstein ring. This ring has a characteristic size
\begin{equation}\label{eq:Mlens}
\theta_E = \sqrt{\frac{4GM_\mathrm{L}}{c^2} (\pi_\mathrm{L} - \varpi)},
\end{equation}
where $G$ is the gravitational constant, $c$ is the speed of light, $M_\mathrm{L}$ is the mass of the lens, $\pi_\mathrm{L}$ is the lens parallax, and $\varpi$ the parallax of the source \citep{Paczynski1986,Wyrzykowski2023}.

The characteristic timescale of a microlensing event, the time in which the source crosses one Einstein radius, is defined as
\begin{equation}
t_\mathrm{E} = \frac{\theta_\mathrm{E}}{\mu_\mathrm{rel}},
\end{equation}
where $\mu_\mathrm{rel}$ is the value of the relative proper motion between source and lens. We note that the duration of an astrometric microlensing event is typically longer than $t_\mathrm{E}$ since astrometric microlensing can be observed at larger impact parameters than photometric microlensing due to the weaker dependence on the impact parameter \citep{Miralda_1996ApJ...470L.113M, Paczynski1986}.

{We define $u_0$ as the angular impact parameter  --- the closest approach between the lens and the source, which occurs at time $t_0$ --- divided by $\theta_\mathrm{E}$. The angular separation between the source and the lens, divided by $\theta_\mathrm{E}$, is thus given by
\begin{equation}
u(t) = \sqrt{\left(\frac{t - t_0}{t_\mathrm{E}} + \delta \tau_1\right)^2 + \left(u_0 + \delta \tau_2\right)^2}.
\end{equation}
$\delta\tau_1$ and $\delta\tau_2$ are the two components of the microlensing parallax displacement vector $(\delta \tau_1, \delta \tau_2)$, which is perpendicular to the line of sight to the lens and source  \citep{Gould_2004ApJ...606..319G}. To express this vector and describe the impact of microlensing parallax, we define a right-handed coordinate system aligned with the source position, where \(\boldsymbol{\hat{n}}\) and \(\boldsymbol{\hat{e}}\) are the unit vectors pointing north and east, respectively. The Sun’s position projected onto this frame is given by
\begin{equation}
    (s_n, s_e) = (\Delta \mathbf{s} \cdot \boldsymbol{\hat{n}}, \Delta \mathbf{s} \cdot \boldsymbol{\hat{e}}),
\end{equation}
where \(\Delta \mathbf{s}\) is the Sun’s positional offset in the geocentric frame. The displacement vector due to microlensing parallax is then expressed as  
\begin{equation}
    (\delta \tau_1, \delta \tau_2) = (\pi_{\mathrm{E}, N}  s_n + \pi_{\mathrm{E}, E}  s_e ,\, -\pi_{\mathrm{E}, E}  s_n + \pi_{\mathrm{E}, N}  s_e).
\end{equation}
\(\pi_{\mathrm{E},N}\) and \(\pi_{\mathrm{E},E}\) are the north and east components of the microlensing parallax vector \(\boldsymbol{\pi}_\mathrm{E}\). \(\boldsymbol{\pi}_\mathrm{E}\) is a two-dimensional vector parallel to the relative proper motion of the lens-source system and can be expressed as 
\begin{equation}
    \pi_\mathrm{E} = |\boldsymbol{\pi}_\mathrm{E}| = \sqrt{\pi_{\mathrm{E},N}^2 + \pi_{\mathrm{E},E}^2}.
\end{equation}
This, in turn, allows us to determine the lens parallax using  
\begin{equation}\label{eq:pi_L}
    \pi_\mathrm{L}=\theta_\mathrm{E} \pi_\mathrm{E} + \varpi,
\end{equation}
where \(\varpi\) is the parallax of the source.}

Microlensing events generally cause the formation of two images of the source. However, since the separation between these images is typically of the order of milliarcseconds and is impossible to resolve, as in observations with \textit{Gaia}, only the centroid of light from the combined images is measured. Its position relative to the lens is described by
\begin{equation}
\theta_C = \frac{u(t)^2 + 3}{u(t)^2 + 2} u(t) \theta_\mathrm{E}.
\end{equation}
The astrometric shift with respect to the unlensed track is given by ( \citealt{Dominik_sahu_2000ApJ...534..213D, Belokurov2002})
\begin{equation}\label{eq:dthetaC}
\delta \theta_C = \frac{u(t)}{u(t)^2 + 2} \theta_\mathrm{E}.
\end{equation}
The maximum shift of the light centroid occurs at $u(t) = \sqrt{2}$. 

In general, the light from the source can blend with a luminous lens, affecting the centroid deviation described by Equation \ref{eq:dthetaC} \citep{Hog1995, Miyamoto1995, Walker1995}. However, in our study, we focused on dark lenses and ignored any additional contribution to the light observed from sources not related to the event. In this scenario, the shift of the light centroid traces out an ellipse in the sky (e.g., \citealt{Dominik_sahu_2000ApJ...534..213D}). The shape and orientation of this ellipse could, in principle, provide information about the lensing parameters, including the mass and distance of the lens. By fitting the theoretical model to the observed astrometric shifts, it is possible to derive the properties of the lensing object \citep{Belokurov2002}.

\subsection{Generating mock \textit{Gaia} observations}\label{sec:mock_obs}

We generated \textit{Gaia} {mock datasets} corresponding to \textit{Gaia} DR4 run from 2014.5 to 2020 with \texttt{astromet}. To obtain the \textit{Gaia} scanning laws, the scanning angles $\varphi_\mathrm{obs}$ as a function of the observation time $t_\mathrm{obs}$, we used the Gaia Observing Schedule Tool (GOST)\footnote{\href{https://gaia.esac.esa.int/gost/}{https://gaia.esac.esa.int/gost/}}. We set the reference epoch at $t_\mathrm{ref} = 2017.5$ and assumed that there is no blending in the observations (neither from the lens nor from unrelated sources). The position {and motion} of a source on the celestial sphere at $t_\mathrm{ref}$ is defined by its Right Ascension $\alpha_0$, Declination $\delta_0$, proper motion in Right Ascension $\mu_{\alpha^*}$,\footnote{We used a “$*$” symbol for {motions corrected for $\cos\delta$; $\mu_{\alpha^*} = \mu_{\alpha}\cos\delta$.}} proper motion in Declination $\mu_\delta$, and parallax $\varpi$. We generated six microlensing {mock datasets}. Three data sets correspond to three different baseline G-band magnitudes $G_0=14$, 16.5, and 19 mag, assuming the source position $(\alpha_0,\delta_0)=(6.5^\circ,-47.3^\circ)$. Three data sets correspond to three different source positions $(\alpha_0, \delta_0)=(6.5^\circ, -47.3^\circ),\,(277.8^\circ, -68.3^\circ),\,(287.9^\circ,45.9^\circ)$ as shown in Table \ref{tab:micro_params}, assuming $G_0=14$, which are characterized by the number of \textit{Gaia} visits $N_\mathrm{v} = 281$, 209, and 91, respectively.\footnote{{These $N_\mathrm{v}$ values correspond to regions most frequently visited by \textit{Gaia} (281 visits), regions with an approximately average visit number (91 visits), and intermediate regions (209 visits). We determine $N_\mathrm{v}$ using the \texttt{scanninglaw} Python package (\href{https://github.com/gaiaverse/scanninglaw}{https://github.com/gaiaverse/scanninglaw}), as this information is not available in GOST. However, since this package does not provide data for \textit{Gaia} DR4, we used $N_\mathrm{v}$ for \textit{Gaia} DR3.}}. We also generate one additional data set with $t_0\in[2010,2014]\,$yr, $G_0=14$, $(\alpha_0,\delta_0)=(6.5^\circ,-47.3^\circ)$, meaning that the peak of the microlensing signal was located outside of the DR4 time span. To generate mock \textit{Gaia} observations of microlensing events, we used five single-source parameters and six microlensing parameters. The single source parameters are: $\alpha_0$, $\delta_0$, $\mu_{\alpha^*}$, $\mu_\delta$, and $\varpi$. The microlensing parameters are $u_0$, $\theta_\mathrm{E}$, $t_0$, $t_\mathrm{E}$, $\pi_\mathrm{EE}$, and $\pi_\mathrm{EN}$. The specific ranges of microlensing parameters that we used to generate mock data are shown in Table \ref{tab:micro_params}. For each {mock dataset}, we simulate 50,000 microlensing events.

\begin{table}
    \caption{Microlensing parameter ranges.}
    \label{tab:micro_params}
    \centering
    \begin{tabular}{l l}
        \hline\hline
        Parameter & Range \\
        \hline\hline
        $\alpha_0\,\mathrm{[^\circ]}$ & $\{6.5, 277.8,   287.9\}$ \\
        $\delta_0\,\mathrm{[^\circ]}$ & $\{-47.3, -68.3, 45.9\}$ \\
        $\mu_{\alpha^*}\,\mathrm{[mas/yr]}$ & $[-40, 40]$ \\
        $\mu_\delta\,\mathrm{[mas/yr]}$  & $[-40, 40]$ \\
        $\varpi\,\mathrm{[mas]}$ & $[0.01,2]$ \\
        $u_0$ & $[-5,5]$ \\
        $\theta_\mathrm{E}\,\mathrm{[mas]}$ & $[0.01,10]$ \\
        $t_0\,\mathrm{[yr]}$ & $[2014.5, 2020]$ or $[2010, 2024]$ \\
        $t_\mathrm{E}\,\mathrm{[yr]}$ & $[0.01, 2]$ \\
        $\pi_\mathrm{EE}$ & $[-1, 1]$ \\
        $\pi_\mathrm{EN}$ & $[-1, 1]$ \\
        \hline
    \end{tabular}
    \tablefoot{Microlensing parameter ranges used to generate mock \textit{Gaia} astrometric observations. We consider three different sky positions observed with \textit{Gaia}, while the other parameters are generated from random uniform distributions within the specified ranges.}
\end{table}

\begin{figure}
	\centering
	\begin{minipage}[b]{\linewidth}
		\includegraphics[width=\textwidth]{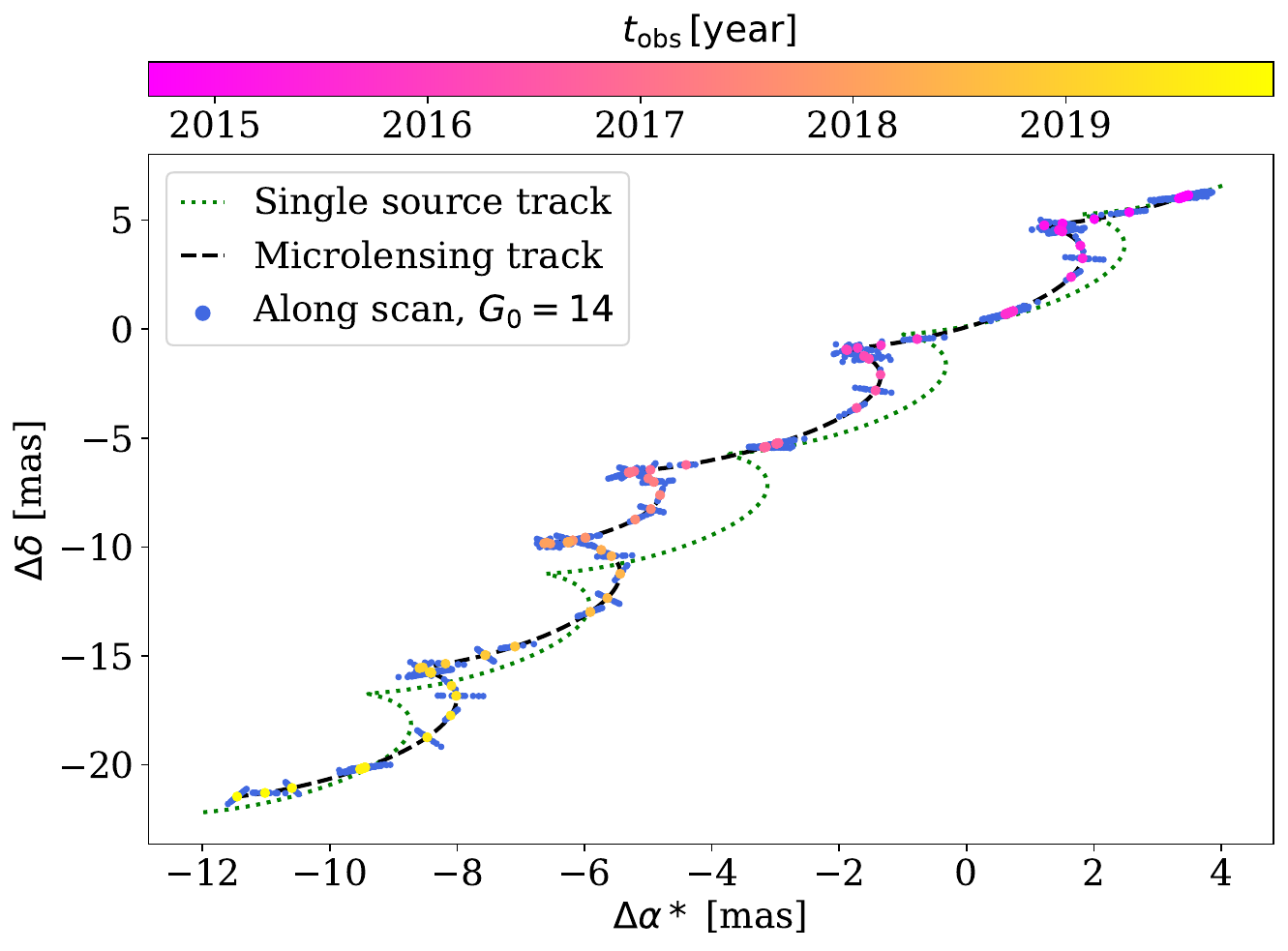}
	\end{minipage}
        \begin{minipage}[b]{\linewidth}
		\includegraphics[width=\textwidth]{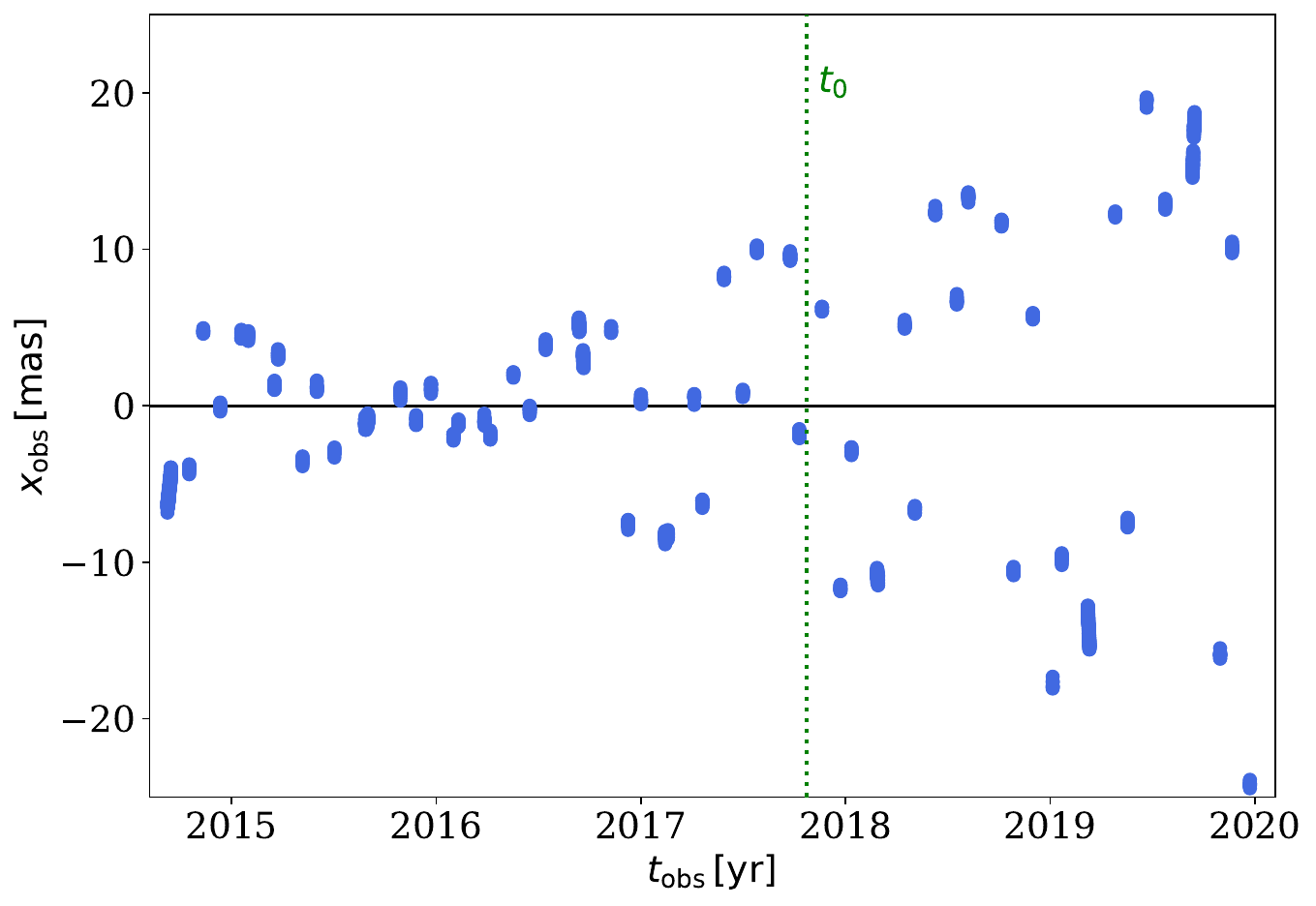}
	\end{minipage}
        \begin{minipage}[b]{\linewidth}
		\includegraphics[width=\textwidth]{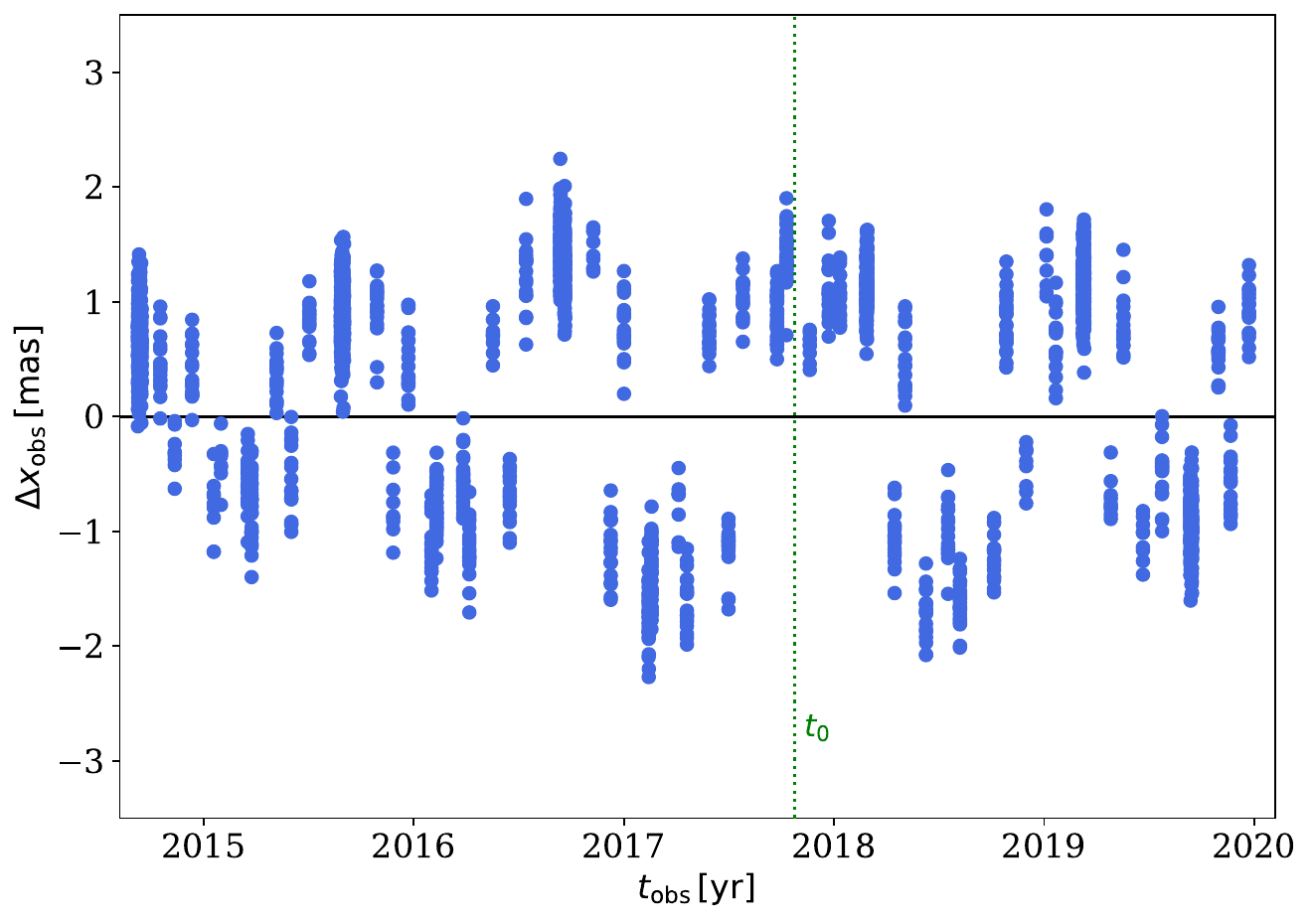}
	\end{minipage}
 
	\caption{{Upper panel}: Tracks of the light centroid
for a single source event (green, dotted line) and a microlensing event with the same single source parameters (black,
dashed line). The blue circle symbols correspond to \textit{Gaia} along scan measurements at different $t_\mathrm{obs}$ denoted by the coloured circle symbols. {Middle panel}: \textit{Gaia} along scan measurements $x_\mathrm{obs}$ as a function of $t_\mathrm{obs}$ for the microlensing event.  {Lower panel}: Residuals between the simulated data with microlensing vs model with no microlensing ($\Delta x_\mathrm{obs}$) as a function of $t_\mathrm{obs}$. The green vertical dashed lines correspond to $t_0$. For all the panels, we assume the following parameters: $\alpha_0 = 6.5{^\circ}$, $\delta_0 = -47.3{^\circ}$, $\mu_{\mathrm{\alpha}^*} = -2.8{\text{ mas/yr}}$, $\mu_\mathrm{\delta} = -5.5{\text{ mas/yr}}$, $\varpi = 1{\text{ mas}}$, $u_0=-0.6$, $\theta_\mathrm{E}=5{\text{ mas}}$, $t_\mathrm{E} = 100{\text{ days}}$, $t_0 = 2017.8$, $\pi_\mathrm{EE} = -0.1$, $\pi_\mathrm{EN} = -0.1$, and $G_0=14$.}
	\label{fig:tracks}
\end{figure}

In Figure \ref{fig:tracks} (top panel), we show the light centroid tracks generated for a single source event (green, dotted line) and a microlensing event with the same single source parameters and microlensing parameters $u_0=-0.6$, $\theta_\mathrm{E}=5$, $t_\mathrm{E} = 100$, $t_0 = 2017.8$, $\pi_\mathrm{EE} = -0.1$, $\pi_\mathrm{EN} = -0.1$ (black, dashed line). The coloured circles correspond to $t_\mathrm{obs}$, while the blue circle symbols indicate \textit{Gaia} along scan measurements. We used the light centroid tracks to calculate the deviations along the scanning direction $x_\mathrm{obs}$, which are shown in Figure \ref{fig:tracks} (middle panel). {Each generated track includes magnitude measurements, which are used to obtain the astrometric measurement errors $x_\mathrm{err}$  by scattering the astrometric positions according to a Gaussian distribution}. In Figure \ref{fig:tracks} (lower panel), we show residuals $\Delta x_\mathrm{obs}$, {calculated as the difference between $x_\mathrm{obs}$ and the modelled deviations along the scanning direction using parameters from fitting a single source model to $x_\mathrm{obs}$}. Hence, $\Delta x_\mathrm{obs}$ is the deviation from the single source track, which offers a direct indication of the microlensing event and provides a way of estimating $\theta_\mathrm{E}$, $t_0$, and $t_\mathrm{E}$. We elaborate on this in more detail in Section \ref{sec:game_filter}.

Other types of astrometric events can produce light centroid tracks similar to those of microlensing events, which can lead to event misclassification. To assess the rate of such false-positive microlensing detections, we generated a mock \textit{Gaia} dataset of 50,000 binary system events using a procedure similar to that used for microlensing events. The required parameters are five single-source parameters and nine binary system parameters. The parameters of binary systems are the orbital period $P$, the orbital semimajor axis $a$, the orbital eccentricity $e$, the ratio between masses of both stars $q$, the ratio between luminosities of both stars $l$, the polar viewing angle
$\theta_\mathrm{v}$, the azimuthal viewing angle $\phi_\mathrm{v}$, the planar projection angle $\omega_\mathrm{v}$, and time of periapse passage $t_{\text{peri}}$ \citep{Penoyre_2022MNRAS.513.2437P}. The specific ranges of the binary system parameters used to generate mock data are gathered in Table \ref{tab:binary_params}.

 \begin{table}
    \caption{Binary system parameter ranges.}
    \label{tab:binary_params}
    \centering
    \begin{tabular}{l l}
        \hline\hline
        Parameter & Range \\
        \hline\hline
        $P\,\mathrm{[yr]}$ & $[0.01, 5]$ \\
        $a\,\mathrm{[AU]}$ & $[0.1, 10]$ \\
        $e$ & $[0, 0.9]$ \\
        $q$ & $[0.01, 1]$ \\
        $l$ & $[0.01, 1]$ \\
        $t_{\mathrm{peri}}\,\mathrm{[yr]}$ & 2016 \\
        $\theta_\mathrm{v}\,\mathrm{[^\circ]}$ & $[-180, 180]$ \\
        $\phi_\mathrm{v}\,\mathrm{[^\circ]}$ & $[0, 360]$ \\
        $\omega_\mathrm{v}\,\mathrm{[^\circ]}$ & $[0, 360]$ \\
        \hline
    \end{tabular}
    \tablefoot{Binary system parameter ranges used to generate mock \textit{Gaia} astrometric observations. We consider a specific $t_{\mathrm{peri}}$, while the other parameters are generated from random uniform distributions limited by the specific parameter range. The single source parameter ranges and distributions are the same as for microlensing events (see Table \ref{tab:micro_params}).}
\end{table}

\begin{table*}
    \caption{Mock datasets.}
    \label{tab:event_params}
    \centering
    \begin{tabular}{l c c c c c c c}
        \hline\hline
        Dataset name & Event type & $G_0$ & $N_\mathrm{v}$ & $t_0$ range & $P_\mathrm{rec}$ & $P_{20}\,\mathrm{[\%]}$ & $P_{10}\,\mathrm{[\%]}$ \\
        \hline\hline
        \texttt{lens\_G14\_N281\_DR4} & Microlensing & 14 & 281 & [2014.5, 2020] & 61.0\% & 55.0\% & 38.9\% \\
        \hdashline
        \texttt{lens\_G16.5\_N281\_DR4} & Microlensing & 16.5 & 281 & [2014.5, 2020] & 43.1\% & 28.9\% & 15.2\% \\
        \texttt{lens\_G19\_N281\_DR4} & Microlensing & 19 & 281 & [2014.5, 2020] & 16.3\% & 3.4\% & 0.5\% \\
        \hdashline
        \texttt{lens\_G14\_N209\_DR4} & Microlensing & 14 & 209 & [2014.5, 2020] & 58.9\% & 55.0\% & 38.9\% \\
        \texttt{lens\_G14\_N91\_DR4} & Microlensing & 14 & 91 & [2014.5, 2020] & 59.3\% & 50.6\% & 33.9\% \\
        \hdashline
        \texttt{lens\_G14\_N281\_extended}  & Microlensing & 14 & 281 & [2010, 2024] & 46.9\% & 34.5\% & 23.4\% \\
        \hdashline
        \texttt{binary\_G14\_N281\_DR4} & Binary & 14 & 281 & [2014.5, 2020] & 5\% & 0.0\% & 0.0\% \\
        \hline
    \end{tabular}
    \tablefoot{Mock datasets with their respective parameters and accuracy of parameter estimation. {In the first column, we list the names of all mock datasets, where we adopt the following naming convention. The prefixes “lens” and “binary” indicate a microlensing or binary mock dataset, respectively. The number following “G” refers to the baseline G-band magnitude of the source $G_0$. The number following “N” represents the number of \textit{Gaia} visits $N_\mathrm{v}$ for that particular sky position. The suffixes “DR4” and “extended” denote whether the mock dataset assumes that the time of closest approach $t_0$ falls within the \textit{Gaia} DR4 time span or extends beyond it, respectively.}}
\end{table*}

{In Table \ref{tab:event_params}, we list all the mock datasets. We assumed a uniform distribution of single-source, microlensing, and binary parameters listed in Tables \ref{tab:micro_params} and \ref{tab:binary_params}. This allowed us to systematically evaluate GAME Filter’s performance across a wide range of potential lensing configurations without being restricted to specific astrophysical assumptions or parameter distributions. While a more physically motivated distribution of lens and source properties (e.g., derived from Galactic models; e.g. \citealt{Mao_1996}) would be necessary for a statistical study of microlensing rates, we focused on assessing the filter’s robustness and sensitivity rather than estimating event occurrence rates.}

\subsection{GAME Filter}\label{sec:game_filter}

GAME Filter\footnote{\href{https://github.com/tajjankovic/GAME-Filter/}{https://github.com/tajjankovic/GAME-Filter/}} is a software tool developed to identify microlensing events in the \textit{Gaia} dataset and obtain the parameters of the event that can help derive parameters and properties of the lensing object. The software reads $x_\mathrm{obs}$, $x_\mathrm{err}$, $\Delta x_\mathrm{obs}$, $t_\mathrm{obs}$, and $\varphi_\mathrm{obs}$ from the \textit{Gaia} data files. GAME Filter then calculates $x_\mathrm{fit}$, which is the deviation along $\varphi_\mathrm{obs}$ at $t_\mathrm{obs}$, for specific single source and microlensing parameters. After that, the software minimises the renormalized microlensing unit weight error (MUWE)
\begin{equation}\label{eq:muwe}
\mathrm{MUWE} = \left( \sum_{i=1}^{N} \frac{(x_{\mathrm{obs},i} - x_{\mathrm{fit},i})^2 /x_{\mathrm{err},i}^2}{N-11} \right)^{1/2},
\end{equation}
a scalar parameter which indicates the goodness of the microlensing fit, where $N$ is the number of visits for a specific event. The goal of the minimisation is to identify parameters that bring MUWE as close as possible to its theoretical minimum value of 1. The minimisation process uses the limited memory Broyden-Fletcher-Goldfarb-Shanno (L-BFGS-B) algorithm\footnote{We tested several minimisation methods and found that the L-BFGS-B method produces the most accurate results. This is mainly due to its feature of constraining individual parameters within bounds} to explore the parameter space and determine the optimal single source and microlensing parameters for individual events. Appendix \ref{app:game_filter} provides a more detailed description of the minimisation process.

Following the minimisation process, the minimiser might stop at an incorrect local minimum, failing to find the correct solution. Consequently, we established criteria to determine when an event is recovered. These criteria are based on the value of MUWE after
minimisation $\mathrm{MUWE}_\mathrm{min}$, L2 optimality error $L_\mathrm{opt}$ (see Appendix \ref{app:game_filter}), initial guesses, and the boundaries imposed on individual parameters. {We determined the critical values of $\mathrm{MUWE}_\mathrm{min}$ and $L_\mathrm{opt}$ from their respective histograms as the values where histograms show a sharp decline}. We considered an event to be recovered if the following criteria were met.
\begin{itemize}
    \item $0.9 < \mathrm{MUWE}_\mathrm{min} < 1.1$.
    \item $L_\mathrm{opt} < 0.015$.
    \item The values of $\pi_\mathrm{EE}$, $\pi_\mathrm{EN}$, and $u_0$ differ from the initial guesses.
    \item The values of all parameters are within the boundaries imposed.
\end{itemize}

\section{Results}
\label{sec:results}

To assess the accuracy of GAME filter, we generated mock datasets as described in Section \ref{sec:mock_obs}, compared the recovered parameters to their true values, and investigated degeneracies between parameters (Section \ref{sec:degen}). We considered various scenarios, including different magnitudes (Section \ref{sec:mags}), varying numbers of \textit{Gaia} visits (Section \ref{sec:Nvisit}), detection of events with the peak microlensing signal outside of the \textit{Gaia} observational window (Section \ref{sec:t0_range}), and false positive detections (Section \ref{sec:binary}). 

In Table \ref{tab:event_params}, we list all the {mock datasets} with their abbreviated names and properties. Furthermore, we show the percentages of recovered events $P_\mathrm{rec}$ and the percentages of events with parameter values obtained after minimisation within 10\% (20\%) of their true values, denoted as $P_{10}$ ($P_{20}$). These metrics provide a direct measure of GAME filter's accuracy. Detailed discussions on these results are provided in the following subsections.

The true parameter values used to generate mock observations are: the source's Right Ascension ($\alpha_0$), Declination ($\delta_0$), proper motion in Right Ascension ($\mu_{\alpha^*}$), proper motion in Declination ($\mu_\delta$), and parallax ($\varpi$). The microlensing parameters are: the impact parameter ($u_0$), the Einstein radius ($\theta_\mathrm{E}$), the time of the closest approach between the source and the lens ($t_0$), the event timescale ($t_\mathrm{E}$), and the microlensing parallax components ($\pi_\mathrm{EE}$ and $\pi_\mathrm{EN}$). We refer to the parameter values obtained after minimisation as $\alpha_{0,\mathrm{min}}$, $\delta_{0,\mathrm{min}}$, $\mu_{\alpha^*,\mathrm{min}}$, $\mu_{\delta,\mathrm{min}}$, $\varpi_\mathrm{min}$, $u_\mathrm{0,min}$, $\theta_\mathrm{E,min}$, $t_\mathrm{0,min}$, $t_\mathrm{E,min}$, $\pi_\mathrm{EE,min}$, and $\pi_\mathrm{EN,min}$. In our study, we omit showing the plots for $\alpha_{0,\mathrm{min}}$ and $\delta_{0,\mathrm{min}}$ because we impose tight boundaries on these parameters, and they are always accurately recovered. This is motivated by the fact that the position of the observed region in the sky is always accurately known.

\subsection{Parameter recovery accuracy and degeneracies}\label{sec:degen}

We applied GAME Filter to the \texttt{lens\_G14\_N281\_DR4} {mock dataset} ($G_0=14$, $N_\mathrm{v}=281$, and DR4 observational run) to estimate the accuracy of parameter recovery and investigate degeneracies between parameters. We found that the percentages of recovered events are: $P_\mathrm{rec}=61\%$, $P_{20}=55\%$, and $P_{10}=39\%$ (see Table \ref{tab:event_params}).

\begin{figure}[h]
    \centering
    \begin{minipage}[b]{.99\linewidth}
        \includegraphics[width=\textwidth]{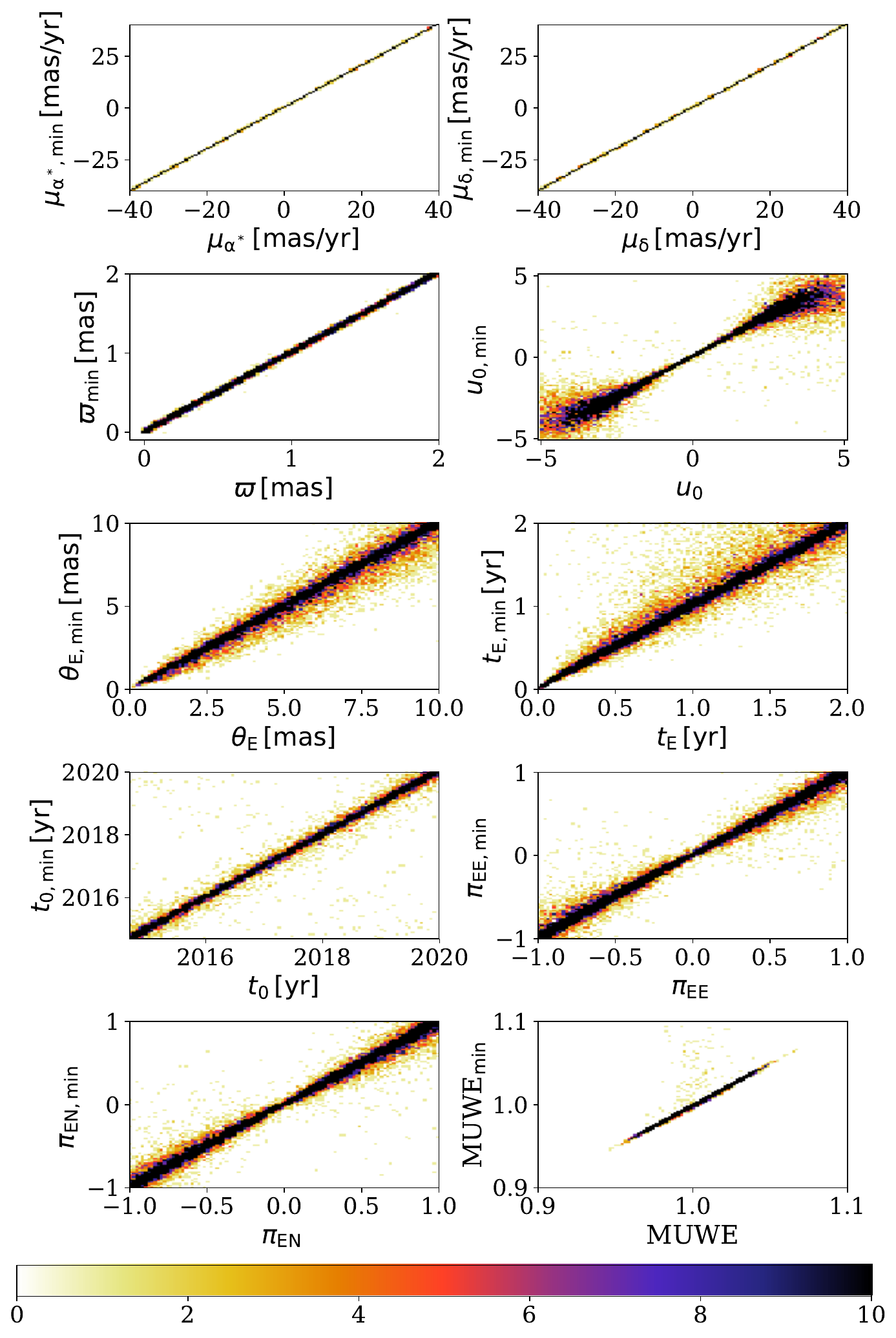}
    \end{minipage}\hfill
    \caption{2D histograms of values of the parameters obtained after minimisation as a function of their true values for the \texttt{lens\_G14\_N281\_DR4} {mock dataset} (see Tables \ref{tab:micro_params} and \ref{tab:event_params}). The colourbar corresponds to the number of events in individual bins.}
    \label{fig:recovered_events}
\end{figure}

In Figure \ref{fig:recovered_events}, we present 2D histograms of the values of the minimised parameters as a function of their true values. A strong linear trend in all cases indicates a good correlation between the true and recovered values. The single source parameter values and $t_0$ are well recovered for the entire parameter range. For high $|u_0|$ values, the scatter around the true values is greater due to the weaker microlensing effect, making the true values more difficult to recover. The duration of a microlensing event is typically longer than $t_\mathrm{E}$. Consequently, the scatter around the true values increases with $t_\mathrm{E}$, as longer events can extend beyond the \textit{Gaia} observational time span, resulting in fewer observations and making $t_\mathrm{E}$ more difficult to recover. Additionally, $\theta_\mathrm{E}$ is more poorly constrained for larger $\theta_\mathrm{E}$, which corresponds to longer distances traversed by the lens, effectively increasing the duration of the microlensing event. Due to this, \textit{Gaia} might observe a smaller fraction of the entire astrometric event.

\begin{figure*}
 	\centering
		\begin{minipage}[b]{.99\linewidth}
		\includegraphics[width=\textwidth]{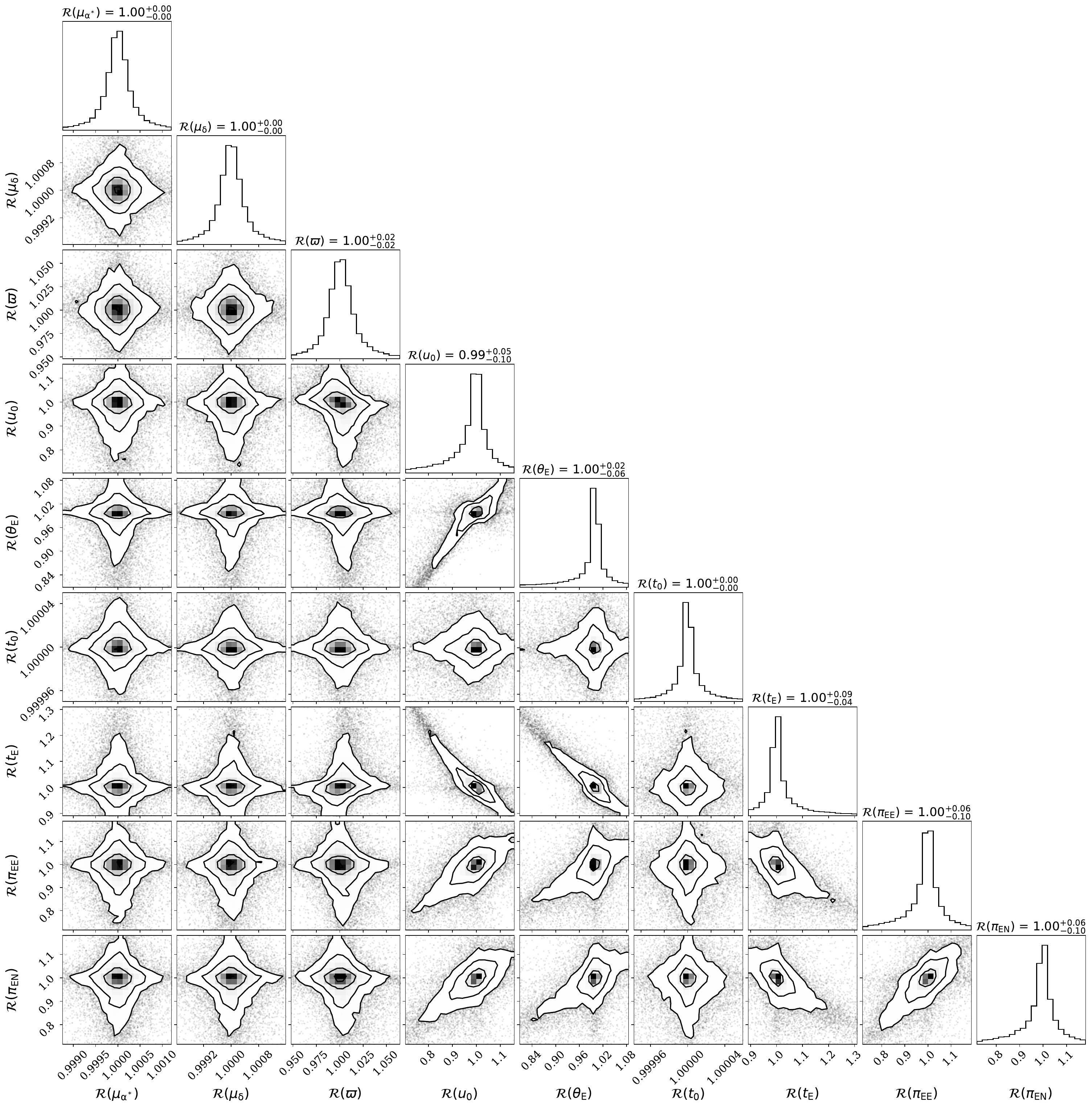}
	\end{minipage}\hfill
 	\caption{Corner plots for ratios $\mathcal{R}$ between true and minimised values of individual parameters for recovered events in the {mock dataset} \texttt{lens\_G14\_N281\_DR4} (see Tables \ref{tab:micro_params} and \ref{tab:event_params}).}
 	\label{fig:corner_plot}
 \end{figure*}

To determine the degeneracies between the parameters, we produced the corner plots shown in Figure \ref{fig:corner_plot}. These plots are calculated for the ratios $\mathcal{R}$ between the true and recovered values of the individual parameters, indicating that values close to 1 imply an accurate recovery of the true parameter values. Due to our method for determining initial guesses (see Appendix \ref{app:game_filter}), the intrinsic relationships between $t_0$, $u_0$, and $\theta_\mathrm{E}$ are reflected in the corner plots. Specifically, the astrometric microlensing signal can exhibit one or two peaks, which influences the initial guess for $t_0$, while the signal amplitude is affected by $\theta_\mathrm{E}$. In photometric microlensing, $\pi_{\mathrm{EE}}$ and $\pi_{\mathrm{EN}}$ exhibit strong degeneracies, resulting in two clusters of data in the corner plot. However, we find that astrometric data break this common degeneracy, as the relationship between $\pi_{\mathrm{EE}}$ and $\pi_{\mathrm{EN}}$ results in a single cluster of data points. This is because $\pi_{\mathrm{EE}}$ and $\pi_{\mathrm{EN}}$ determine the orientation of the ellipse, reducing the expected degeneracy.

\subsection{Parameter recovery for sources with different magnitudes} \label{sec:mags}

We tested GAME Filter on three {mock datasets}, namely \texttt{lens\_G14\_N281\_DR4}, \texttt{lens\_G16.5\_N281\_DR4}, and \texttt{lens\_G19\_N281\_DR4}, which correspond to $G_0=14$, 16.5, and 19, respectively. The results of these tests are summarised in Table \ref{tab:event_params}. As the magnitude increases, noise increases, and the source becomes fainter, the observed signal becomes weaker. Consequently, the number of recovered events decreases and the error of parameter estimation increases, indicated by progressively lower values of $P_\mathrm{rec}$, $P_{20}$, and $P_{10}$ as $G_0$ increases.

\begin{figure}
 	\centering
		\begin{minipage}[b]{.99\linewidth}
		\includegraphics[width=\textwidth]{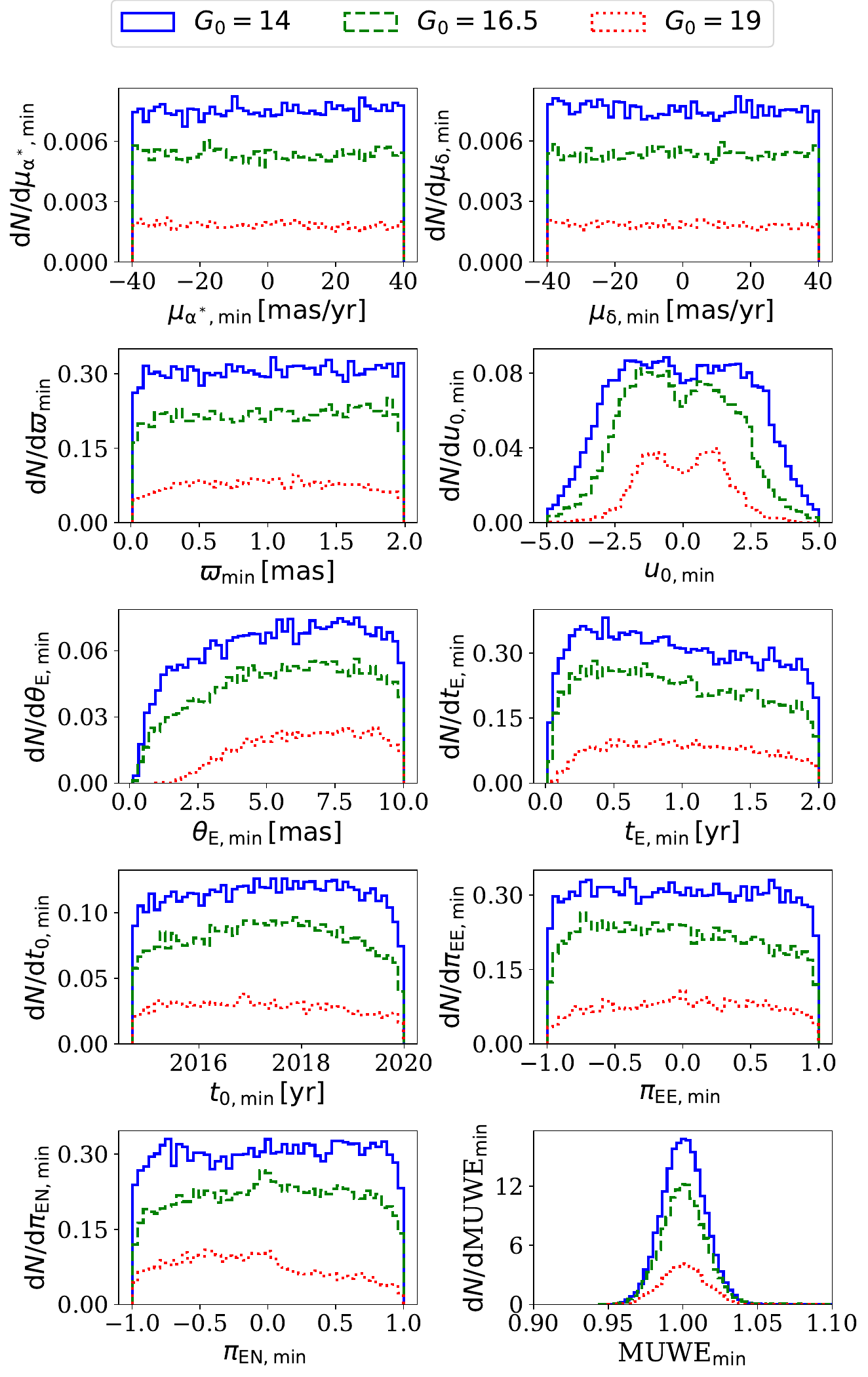}
	\end{minipage}\hfill
 	\caption{Distributions of parameter values after minimisation for recovered events for {mock datasets} \texttt{lens\_G14\_N281\_DR4}, \texttt{lens\_G16.5\_N281\_DR4}, and \texttt{lens\_G19\_N281\_DR4} corresponding to $G_0=$ 14 (blue), 16.5 (green), and 19 (red), respectively. The values in individual bins denote the number of recovered events within the corresponding bin range. The integral of distributions is normalized to $P_\mathrm{rec}$ (see Table \ref{tab:event_params}).}
 	\label{fig:mag_histograms}
 \end{figure}

In Figure \ref{fig:mag_histograms}, we show histograms of the parameter values after minimisation for recovered events for different $G_0$. There are several regions in the parameter space where the number of recovered events is lower: high $|u_0|$, low $\theta_\mathrm{E}$, short $t_\mathrm{E}$, and long $t_\mathrm{E}$. For high $|u_0|$ and low $\theta_\mathrm{E}$, the microlensing signal is inherently weaker. For short and long $t_\mathrm{E}$, \textit{Gaia} might observe only a fraction of the microlensing event: short $t_\mathrm{E}$ due to infrequent \textit{Gaia} observations and long $t_\mathrm{E}$ because the event duration may exceed the \textit{Gaia} DR4 time span. The histograms of $u_0$ show peaks at $u_0 \approx \sqrt{2}$, consistent with Equation \ref{eq:dthetaC}. As the magnitude of the lensed source fades, the number of recovered events decreases, reflecting the increasing difficulty in accurately recovering parameters for fainter sources. The decrease in the number of recovered events is the most significant for regions with a weak microlensing signal. Specifically, there are no recovered events with $\theta_\mathrm{E}\lesssim 2\,$mas or $|u_0|\gtrsim 3$ for $G_0=19$.

{Among all simulated events, $\approx 80\%$  have \( |u_0| > 1 \), meaning that they do not exhibit measurable photometric but only astrometric signals. GAME Filter recovers 55\% of such events for \( G_0 = 14 \), while for events with photometric and astrometric signals (\( |u_0| < 1 \)), the recovery rate increases to 82\%. For fainter source magnitudes with \( G_0 = 16.5 \) and \( G_0 = 19 \), the recovery rates for astrometry-only events decrease to 35.4\% and 10.4\%, respectively.}

\begin{figure}
 	\centering
		\begin{minipage}[b]{.99\linewidth}
		\includegraphics[width=\textwidth]{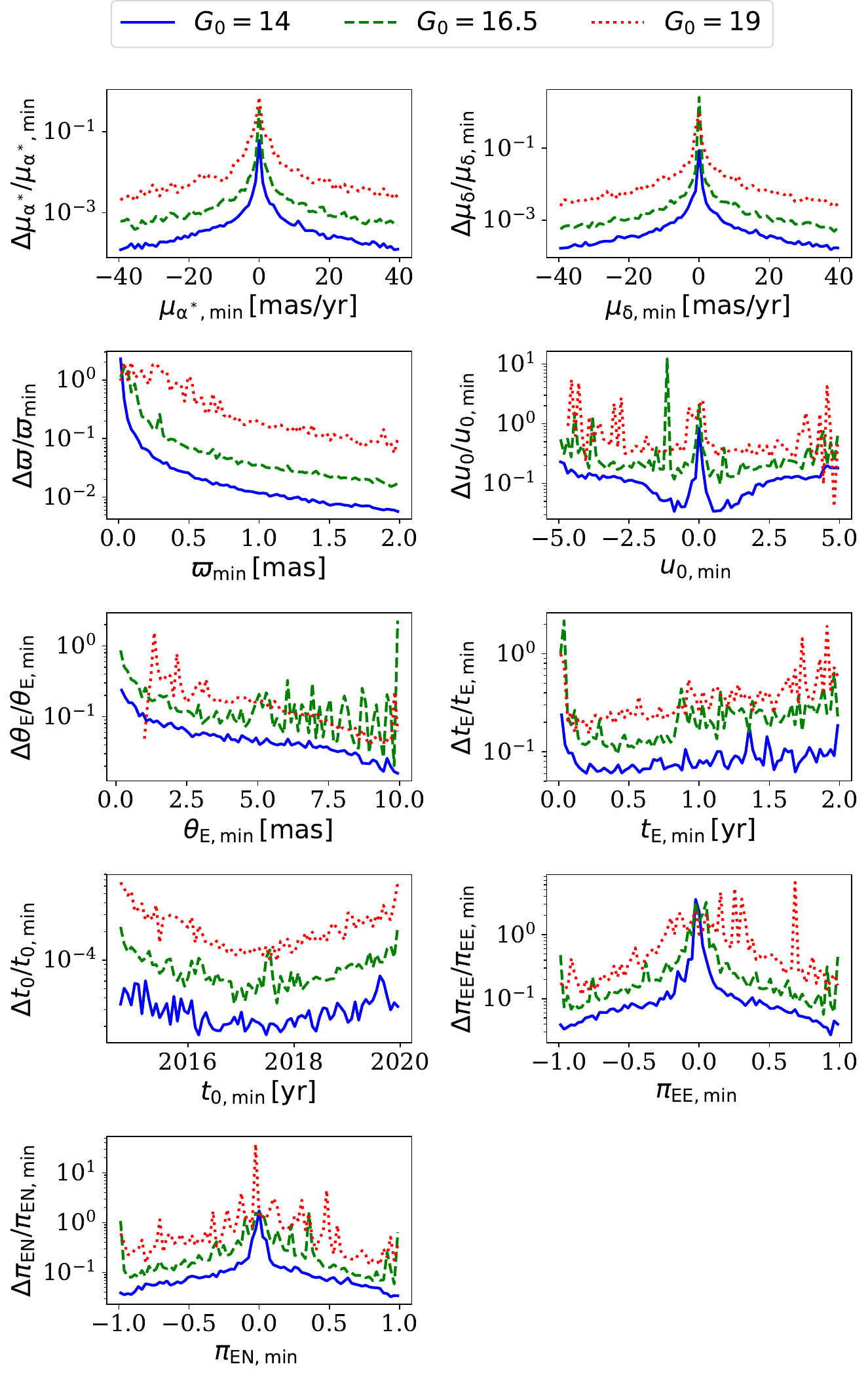}
	\end{minipage}\hfill
 	\caption{Relative errors for recovered events for {mock datasets} \texttt{lens\_G14\_N281\_DR4}, \texttt{lens\_G16.5\_N281\_DR4}, and \texttt{lens\_G19\_N281\_DR4} corresponding to $G_0=$ 14 (blue), 16.5 (green), and 19 (red). The errors are calculated as the difference between the minimised and true parameter values divided by the minimised parameter values. The error values are first binned and then calculated as the mean in each bin.}
 	\label{fig:mag_errors}
 \end{figure}

In Figure \ref{fig:mag_errors}, we show the relative errors for individual parameters after minimisation. The errors are higher for high $|u_0|$, low $\theta_\mathrm{E}$, and short $t_\mathrm{E}$ when the microlensing signal is weaker or more difficult to observe. The relative error of $u_0$ is the lowest at $|u_0| \approx \sqrt{2}$, which corresponds to the maximum microlensing deviation. {When values of proper motion, geometric and microlensing parallaxes, and $u_0$ are small, their relative errors increase drastically due to divisions by values close to 0}. As $G_0$ increases, the errors also increase, indicating that the accuracy of parameter recovery decreases for fainter sources. This change is most pronounced in regions with inherently weaker microlensing signals or fewer observations. However, GAME Filter can still recover parameters even for sources with $G_0 = 19$ with an accuracy $\gtrsim 80$\%, provided that $\theta_\mathrm{E} \gtrsim 2\,$mas. 
 
\subsection{Parameter recovery for events with different numbers of \textit{Gaia} visits}\label{sec:Nvisit}

We tested GAME Filter on {mock datasets} \texttt{lens\_G14\_N281\_DR4}, \texttt{lens\_G14\_N209\_DR4}, and \texttt{lens\_G14\_N91\_DR4}, which correspond to regions with different numbers of \textit{Gaia} visits $N_\mathrm{v}=281$, 209, and 91. The results are summarised in Table \ref{tab:event_params}, showing that the changes in $N_\mathrm{v}$ do not significantly affect $P_\mathrm{rec}$. Additionally,  $P_{20}$ and $P_{10}$ remain unaffected when $N_\mathrm{v}$ decreases to at least 209. When $N_\mathrm{v}$ is further reduced to 91, $P_{20}$ and $P_{10}$ show only a $\approx5$\% change.

\begin{figure}
 	\centering
		\begin{minipage}[b]{.99\linewidth}
		\includegraphics[width=\textwidth]{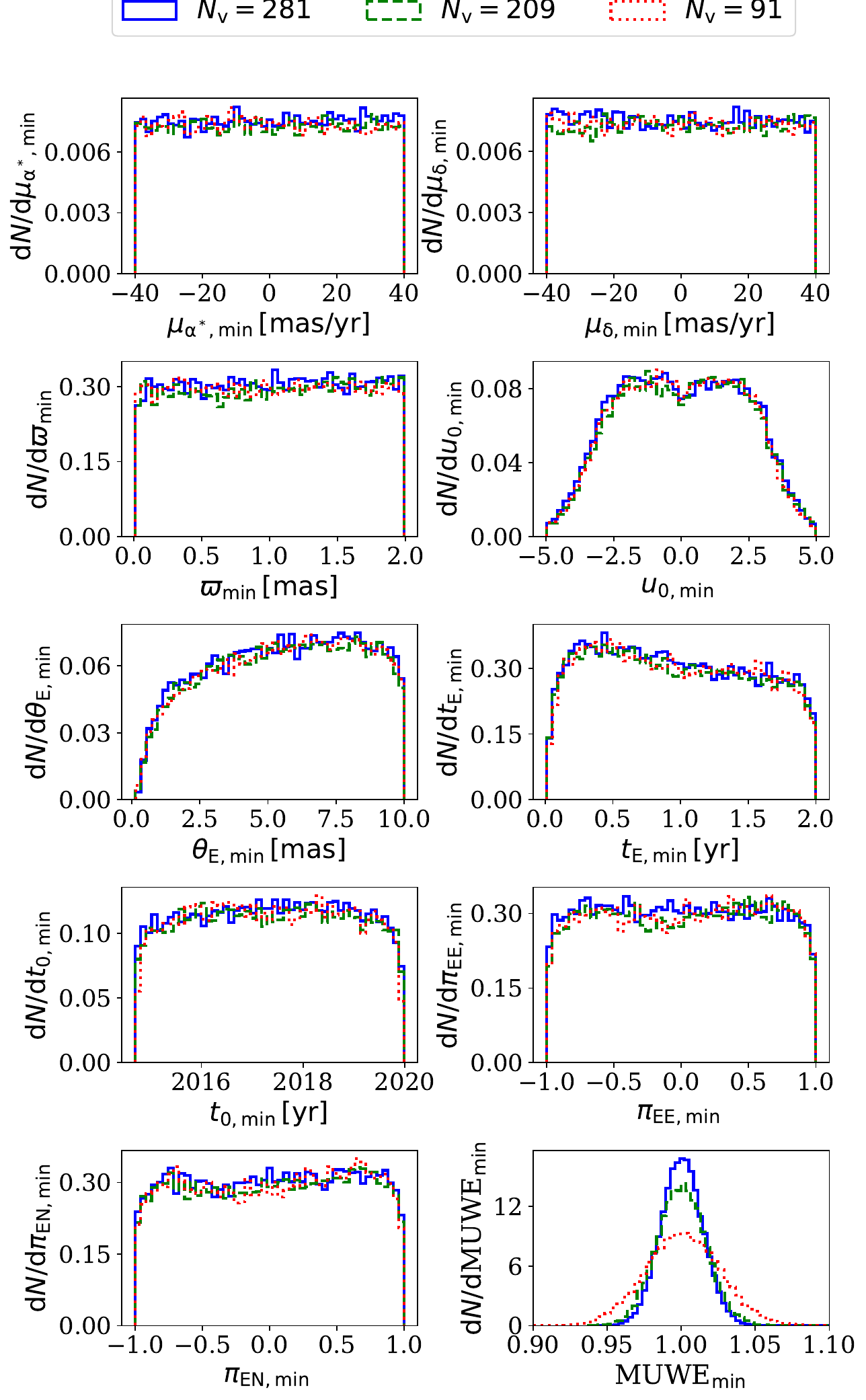}
	\end{minipage}\hfill
 	\caption{Distributions of parameter values after minimisation for recovered events for {mock datasets} \texttt{lens\_G14\_N281\_DR4}, \texttt{lens\_G14\_N209\_DR4}, and \texttt{lens\_G14\_N91\_DR4} corresponding to $N_\mathrm{v}= 281$ (blue), 209 (green), and 91 (red). The values in individual bins denote the number of recovered events within the corresponding bin range. The integral of distributions is normalized to $P_\mathrm{rec}$ (see Table \ref{tab:event_params}).}
 	\label{fig:obs_histograms}
 \end{figure}

In Figure \ref{fig:obs_histograms}, we show histograms of the parameter values after minimisation for recovered events for different $N_\mathrm{v}$. These histograms indicate that $N_\mathrm{v}$ does not significantly affect the recovery of parameter values, as long as $N_\mathrm{v}\gtrsim 90$. The most noticeable difference is in the values of $\mathrm{MUWE}_\mathrm{min}$, which spread more broadly as $N_\mathrm{v}$ decreases.

We also calculated the relative errors for individual parameters after minimisation and found that the errors increase slightly with decreasing $N_\mathrm{v}$. Since the impact is not significant, this indicates that GAME Filter performs robustly even with fewer observations. However, a higher number of observations generally improves the accuracy of parameter recovery.

\subsection{Parameter recovery for events with the maximum microlensing signal beyond \textit{Gaia} observational run}\label{sec:t0_range}

We tested GAME Filter on {mock datasets} \texttt{lens\_G14\_N281\_DR4} and \texttt{lens\_G14\_N281\_extend}, which correspond to {mock dataset} within \textit{Gaia} DR4 with $t_0\in[2014.5, 2020]\,$yr and {mock dataset} extending beyond DR4 with $t_0\in[2010, 2024]\,$yr, respectively. The results of these tests are summarised in Table \ref{tab:event_params}. We see that increasing the $t_0$ range decreases $P_\mathrm{rec}$,  $P_{20}$ and $P_{10}$ by $\approx 15\%$, $\approx 20\%$, and $15\%$, respectively.

\begin{figure}
 	\centering
		\begin{minipage}[b]{.99\linewidth}
		\includegraphics[width=\textwidth]{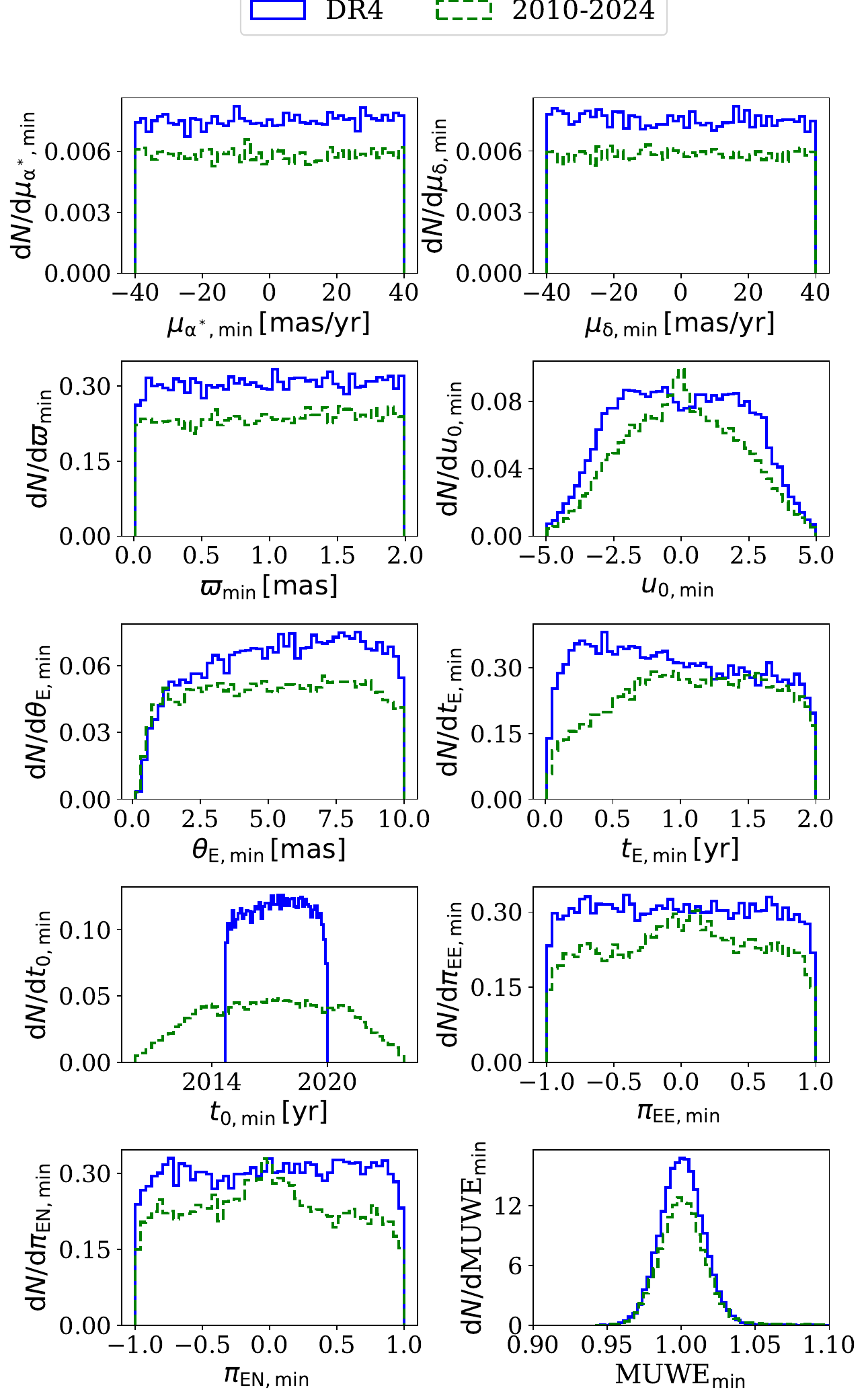}
	\end{minipage}\hfill
 	\caption{Distributions of parameter values after minimisation for recovered events for {mock datasets} \texttt{lens\_G14\_N281\_DR4} and \texttt{lens\_G14\_N209\_extended}, corresponding to $t_0\in[2014.5, 2020]\,$yr and $t_0\in[2010, 2024]\,$yr, respectively. The values in individual bins denote the number of recovered events within the corresponding bin range. The integral of distributions is normalized to $P_\mathrm{rec}$ (see Table \ref{tab:event_params}).}
 	\label{fig:t0_histograms}
 \end{figure}

In Figure \ref{fig:t0_histograms}, we show histograms of the parameter values after minimisation for recovered events for different ranges of $t_0$. As $t_0$ extends beyond the DR4 run, $P_\mathrm{rec}$ decreases. This is because a significant part of the microlensing event can fall outside the \textit{Gaia} observation time span, reducing the amount of available data and making parameter recovery more challenging. This effect is most apparent for events with a short $t_\mathrm{E}$. Furthermore, the histogram of $u_0$ transitions from a bimodal distribution to a single peak as the $t_0$ range increases. There are also fewer recovered events for large $\theta_\mathrm{E}$, as larger $\theta_\mathrm{E}$ corresponds to longer lens paths and event durations, which means that a smaller portion of the event is observed if $t_0$ is outside the DR4 time span. We also see fewer recovered events for microlensing parallaxes close to -1 and 1{, since these events have more variations, or “wiggles” in the observed astrometric signal, which makes the parameter recovery more challenging}. We discuss the physical mechanism for these differences in Section \ref{sec:discussion_accuracy}.

\begin{figure}
 	\centering
		\begin{minipage}[b]{.99\linewidth}
		\includegraphics[width=\textwidth]{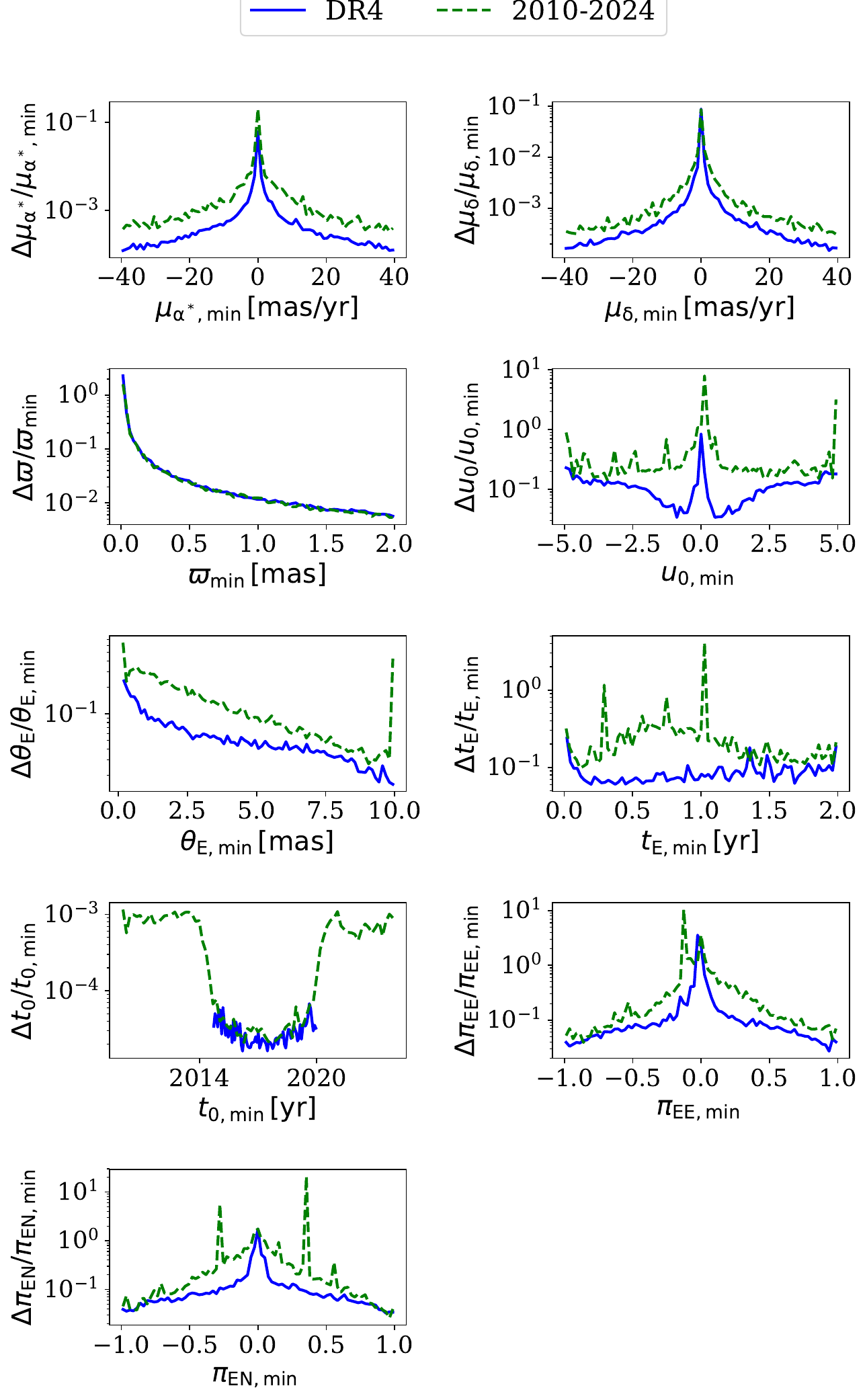}
	\end{minipage}\hfill
 	\caption{Relative errors for recovered events for {mock datasets} \texttt{lens\_G14\_N281\_DR4} and \texttt{lens\_G14\_N209\_extended}, corresponding to $t_0\in[2014.5, 2020]\,$yr and $t_0\in[2010, 2024]\,$yr, respectively. The errors are calculated as the difference between minimised and true parameter values divided by the minimised parameter values. The error values are first binned and then calculated as the mean in each bin.}
 	\label{fig:t0_errors}
 \end{figure}

In Figure \ref{fig:t0_errors}, we show the relative errors for individual parameters after minimisation. The errors are generally higher for all parameters when the $t_0$ range extends beyond the DR4 run,  indicating that the timing of the microlensing event relative to the \textit{Gaia} observational run affects parameter recovery accuracy. This effect is more pronounced for events with short $t_\mathrm{E}$, which are less likely to be fully observed if $t_0$ is outside the DR4 time span.

\subsection{Parameter recovery for the false positive detections}\label{sec:binary}

To evaluate the performance of GAME Filter in distinguishing microlensing events from other astrometric phenomena, we tested it on the \texttt{binary\_G14\_N281\_DR4} {mock dataset}, which corresponds to binary system events. The results of these tests are summarised in Table \ref{tab:event_params}. We determined that the percentage of false positive identifications is very low, with $P_\mathrm{rec}=5\%$ and $P_{20}=P_{10}=0\%$.

\begin{figure}
 	\centering
		\begin{minipage}[b]{.99\linewidth}
		\includegraphics[width=\textwidth]{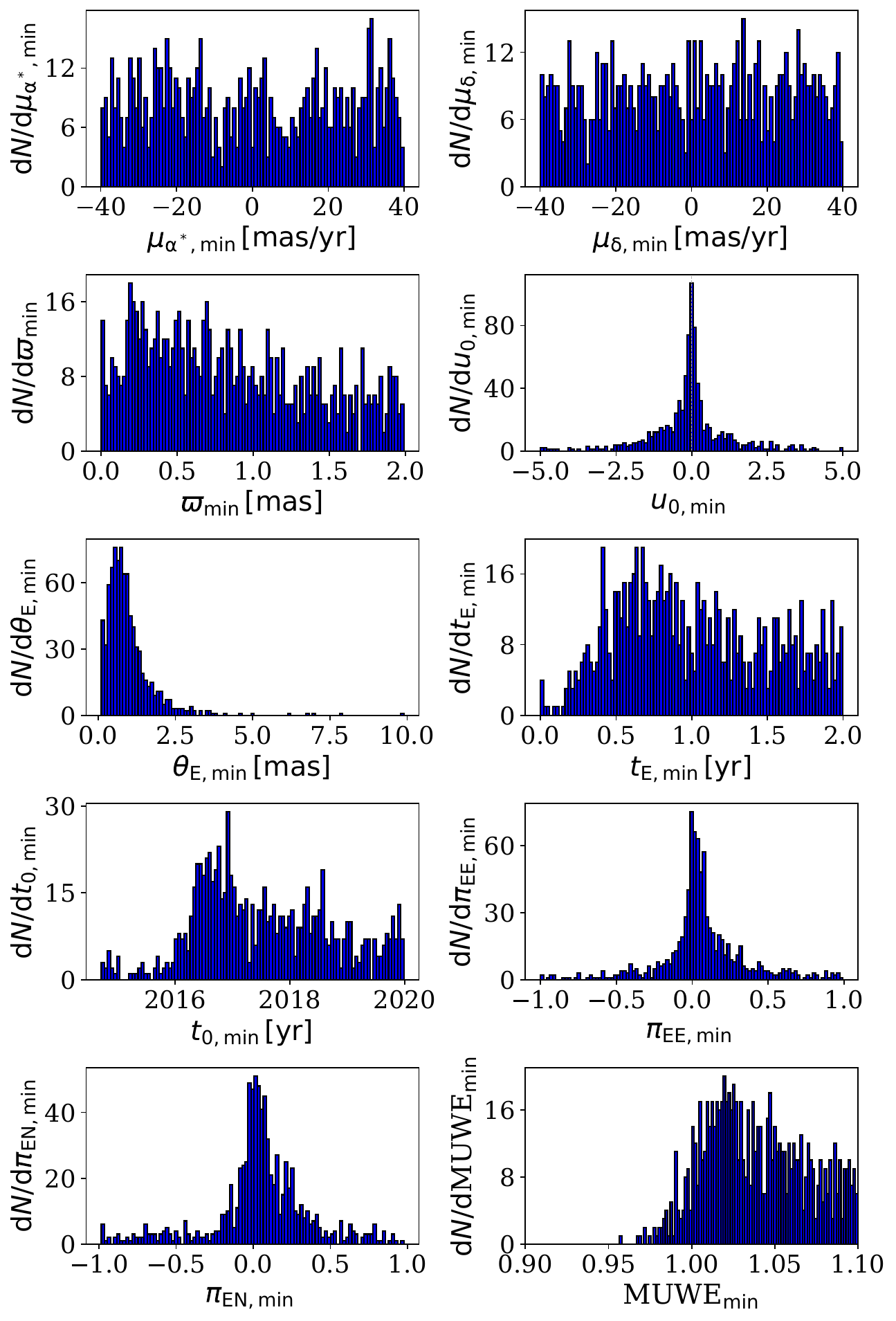}
	\end{minipage}\hfill
 	\caption{Distributions of parameter values after minimisation for binary system events in the \texttt{binary\_G14\_N281\_DR4} {mock dataset} (see Tables \ref{tab:binary_params} and \ref{tab:event_params}). The values in individual bins denote the number of recovered events within the corresponding bin range. The integral of distributions is normalized to $P_\mathrm{rec}$ (see Table \ref{tab:event_params}).}
 	\label{fig:binary_histograms}
 \end{figure}

In Figure \ref{fig:binary_histograms}, we show histograms of the parameter values after minimisation for recovered binary events. Histograms indicate that false positive events typically have low values of $\theta_E$ and high values of $|u_0|$, suggesting that they are interpreted as weak microlensing events. These results demonstrate that while GAME Filter is highly effective in identifying microlensing events, a small percentage of binary events with a weak astrometric signal may still be misinterpreted as microlensing events.

\section{Discussion} \label{sec:discussion}

\subsection{Parameter recovery accuracy}\label{sec:discussion_accuracy}

Given the high number of recovered events and low parameter estimation errors, we find that GAME filter tends to perform with a high level of accuracy. However, there are several challenges related to precisely recovering parameters: i) the effect of microlensing parameters on the microlensing ellipse, ii) the time of closest approach between the lens and the source $t_0$, iii) source magnitude $G_0$, iv) the number of \textit{Gaia} visits $N_\mathrm{v}$,  and v) the probability of false positive detections.

i) The parameters $u_0$ and $\theta_\mathrm{E}$ define the shape and size of this ellipse. A larger $\theta_\mathrm{E}$ results in a proportionally larger ellipse, which increases both the magnitude of the astrometric shift and the duration of the event. When $u_0$ is close to zero, the ellipse becomes more elongated, resembling a straight line, while an increase in $u_0$ makes the ellipse more circular. The strongest microlensing signal occurs at $u_0 = \sqrt{2} \theta_\mathrm{E}$ and $t_0$, where the ellipse reaches its maximum size. This is reflected in the percentage of events recovered for specific regions in the parameter space (see Figures \ref{fig:mag_histograms}, \ref{fig:obs_histograms}, and \ref{fig:t0_histograms}). For high $|u_0|$ and low $\theta_\mathrm{E}$, the microlensing signal is weaker, leading to a decrease in the number of recovered events. Furthermore, high values of $\theta_\mathrm{E}$ can cause a decrease in the number of recovered events due to the longer lens path and duration of the event. This may result in a significant part of the event falling outside the \textit{Gaia} DR4 observational time span, making parameter estimation more challenging. This effect is analogous for long $t_\mathrm{E}$.

\textit{Gaia} observations along the microlensing ellipse are not evenly spaced in distance. For small $u_0$, the observations are more evenly distributed on the sky, whereas for large $u_0$, there are fewer observations near the maximum of the microlensing deviation. Due to this, \textit{Gaia} predominantly observes the astrometric deviation along the edges of the ellipse rather than at its maximum when $t_0$ is outside of DR4, increasing the difficulty of estimating the parameters for high $u_0$. This effect is seen in Figure  \ref{fig:t0_histograms}, where the histogram of $u_0$ changes from a bimodal to a single peak distribution as the range of $t_0$ extends beyond the \textit{Gaia} observational window.

\begin{figure}
	\centering
        \begin{minipage}[b]{\linewidth}
		\includegraphics[width=\textwidth]{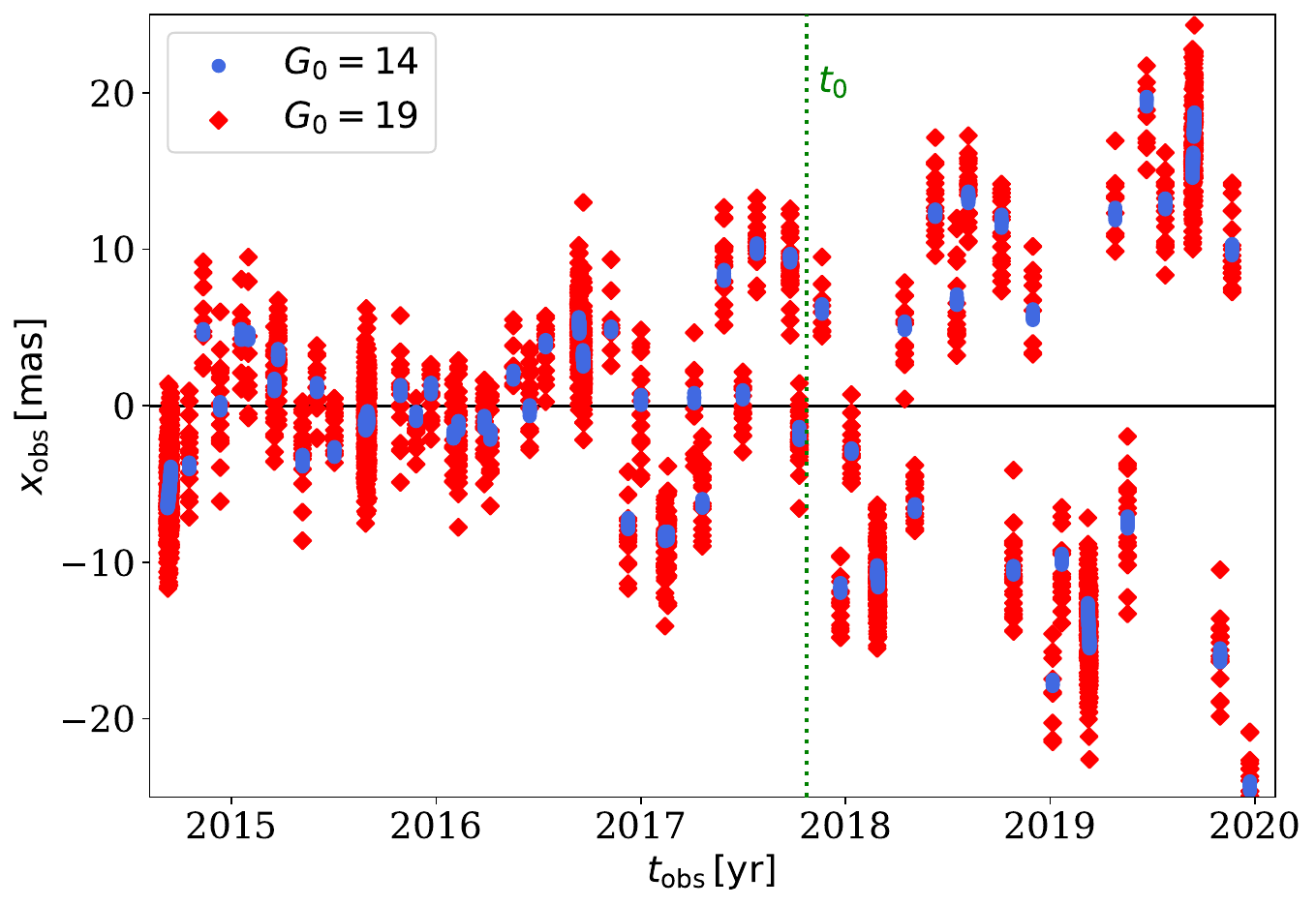}
	\end{minipage}
	\caption{\textit{Gaia} along scan measurements $x_\mathrm{obs}$ as a function of $t_\mathrm{obs}$ for $G_0=14$ (blue circle symbols) and 19 (red diamond symbols). The green vertical dashed lines correspond to $t_0$. We assume the same single source and microlensing parameters as in Figure \ref{fig:tracks}.}
	\label{fig:x_obs_mags}
\end{figure}

The microlensing ellipse is also affected by $\pi_\mathrm{EE}$ and $\pi_\mathrm{EN}$, which determine both the orientation of the ellipse in the sky and the parallax of the lens $\pi_\mathrm{L}$ (see Equation \ref{eq:pi_L}). Higher values of $\pi_\mathrm{EE}$ and $\pi_\mathrm{EN}$ lead to a larger $\pi_\mathrm{E}$ and consequently to a higher $\pi_\mathrm{L}$, assuming a fixed $\varpi$. When the lens is closer to the observer, its parallactic motion introduces more variations in the observed astrometric signal. This added complexity makes parameter estimation more challenging, requiring more measurements to accurately constrain the parameters. This effect is particularly pronounced when $t_0$ lies outside of the \textit{Gaia} observational window, as shown in Figure \ref{fig:t0_histograms}.

ii) Microlensing events can span several years, and many such events have $t_0$ beyond the \textit{Gaia} DR4 observation time span. Consequently, \textit{Gaia} may often miss the maximum of the astrometric microlensing signal. However, we find that when $t_0$ lies outside of \textit{Gaia} DR4, parameters can still be recovered. The percentage of recovered events decreases by $\approx 15\%$ (see Table \ref{tab:event_params}) and the parameter estimation errors for such events are higher (see Figure \ref{fig:t0_errors}). This is most evident for events with short $t_\mathrm{E}$, which are less likely to be observed, as seen in Figure \ref{fig:t0_histograms}.

iii) When simulating \textit{Gaia} mock observations, noise is introduced by scattering the observations along the scanning angle based on the magnitude-dependent error bars, thus reflecting true \textit{Gaia} measurements. Due to this, the scatter in \textit{Gaia}'s along-scan measurements increases with fainter $G_0$ as shown in Figure \ref{fig:x_obs_mags}. This increased scatter makes it more challenging to accurately recover parameters, as seen in Figures \ref{fig:mag_histograms}, \ref{fig:obs_histograms}, and \ref{fig:t0_histograms}. This effect is most pronounced in regions with high $|u_0|$ and low $\theta_\mathrm{E}$, where the microlensing signal is already weak.  Specifically, for $\theta_\mathrm{E} \gtrsim 2.5$ mas, the parameter is recovered with a relative error of $\lesssim 20\%$ even when $G_0 = 19$. However, for $\theta_\mathrm{E} \lesssim 2.5$ mas, the error estimation can reach an order of unity if $G_0 = 19$.

iv) In \textit{Gaia} DR4, different regions in the sky are observed on average $\approx 100$ times. We find that for $N_\mathrm{v}\geq 91$, a decrease in $N_\mathrm{v}$ leads to only a slight increase in relative errors. For example, when $N_\mathrm{v}$ decreases from 281 to 91, the percentage of parameters with their estimated values within 10\% of their true values decreases by $\approx 5\%$ (see Table \ref{tab:event_params}).

v) {\textit{Gaia} observes various types of astrometric phenomena, including single-star motions, binary systems, microlensing, and other transient events.\footnote{ {Unresolved fixed-separation binaries where one star is photometrically variable could represent another source of false positives. However, these can typically be identified through photometric observations. Given Gaia DR4's observational baseline of approximately 5.5 years, most variable sources will exhibit multiple astrometric or photometric peaks within this timeframe, distinguishing them from microlensing events. Nonetheless, transient phenomena such as novae could still potentially mimic microlensing signatures.}} In general, binaries produce periodic astrometric signals that are easily distinguishable from transient microlensing events.} {However, a subset of binaries characterized by long orbital periods and large semi-major axes can produce transient signals mimicking microlensing. We adopted uniform orbital period distributions from 0.01 to 5 years (see Table \ref{tab:binary_params}) and found that $\sim 5$\% of binary events are misclassified as microlensing events for $G_0=14$.\footnote{{For fainter sources, astrometric uncertainties increase, potentially elevating the rate of false positives. However, since the GAME filter preferentially detects high signal-to-noise events at fainter magnitudes, these two effects largely balance out, resulting in a misclassification rate that does not significantly depend on magnitude.}} However, realistic binary populations differ from uniform distributions. Observations of DR3 binaries show that orbital periods typically peak around 500 days \citep{Halbwachs_2023A&A...674A...9H}, while predictions for Gaia DR4 suggest a peak around 2.5–3 years \citep{Badry_2024OJAp....7E.100E}. Thus, our choice likely underrepresents short-period binaries and overrepresents longer-period binaries. Despite this bias, we found that the false positives exhibit uniformly distributed orbital parameters, suggesting that statistical noise rather than particular orbital configurations is the dominant cause of confusion.}

\subsection{Constraining the lens properties}

{To evaluate the sensitivity of the GAME Filter to detect lenses in the Milky Way with varying distances $D_\mathrm{L}$ and masses $M_\mathrm{L}$, we applied the GAME Filter to a new {mock dataset} of $N_\mathrm{all} = 50,000$ microlensing events. For each event, \( M_\mathrm{L} \) is randomly selected from a logarithmically spaced interval between \( 0.1 \) and \( 20\,\mathrm{M_\odot} \), and \( D_\mathrm{L} \) is uniformly sampled from \( 0.1 \) to \( 7.9\,\mathrm{kpc} \). The angle between the microlensing parallax components \( \pi_\mathrm{EE} \) and \( \pi_\mathrm{EN} \) is selected from a uniform distribution between \( 0 \) and \( 2\pi \). Additional parameters are sampled uniformly within their respective ranges (see Table \ref{tab:micro_params}) with the exception of $u_0$ where we changed the range to -3 and 3. Using these values, the microlensing parallax \( \pi_\mathrm{E} \) and the angular Einstein radius \( \theta_\mathrm{E} \) are calculated from \( \pi_\mathrm{S} \), \( M_\mathrm{L} \), and \( D_\mathrm{L} \) (see Equations \ref{eq:Mlens} and \ref{eq:pi_L}). GAME Filter recovers $N_{20\%} = 11,456$ events with parameter values minimised within 20\% of their true values, corresponding to approximately \( 22\% \) of \( N_\mathrm{all} \).}

\begin{figure}
	\centering
		\includegraphics[width=\linewidth]{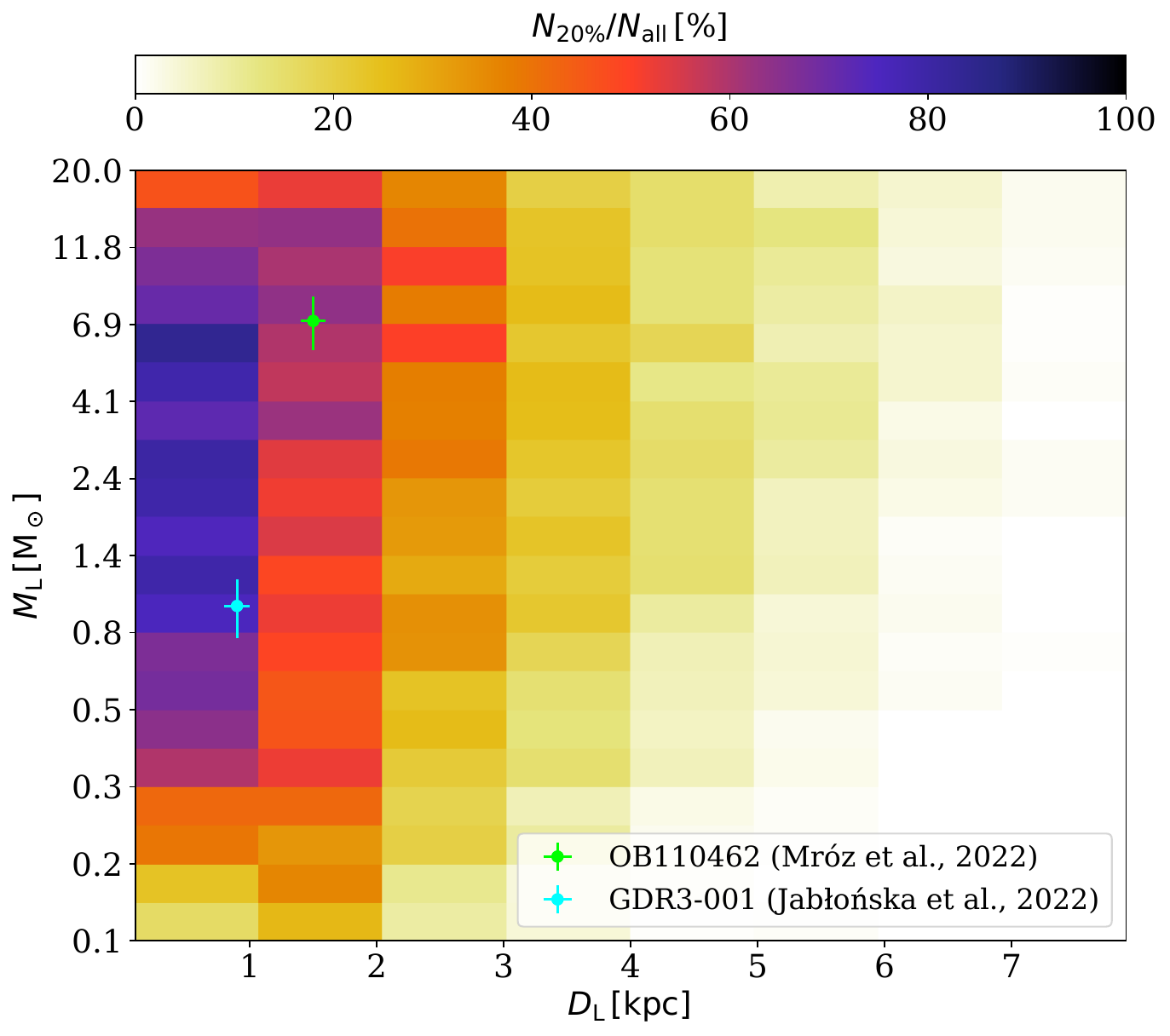}
	\caption{{2D histogram showing the sensitivity of GAME Filter as a function of \(D_\mathrm{L}\) and \(M_\mathrm{L}\). The sensitivity, indicated by the colour bar, is calculated as the ratio \( N_{20\%}/N_\mathrm{all} \). \( M_\mathrm{L} \) is sampled logarithmically between \( 0.1 \) and \( 20\,\mathrm{M_\odot} \), and \( D_\mathrm{L} \) is sampled uniformly between \( 0.1 \) and \( 7.9\,\mathrm{kpc} \). The cyan and green circle simbols with error bars indicate two microlensing events, OB110462 (\( M_\mathrm{L} = 7.1 \pm 1.3\,\mathrm{M_\odot} \), \( D_\mathrm{L} = 1.5 \pm 0.1\,\mathrm{kpc} \); \citealt{Mroz_2022ApJ...937L..24M}) and GDR3-001 (\( M_\mathrm{L} = 1.0 \pm 0.2\,\mathrm{M_\odot} \), \( D_\mathrm{L} = 0.9 \pm 0.1\,\mathrm{kpc} \); \citealt{Jablonska2022}), respectively.}}
	\label{fig:lens}
\end{figure}

{In Figure \ref{fig:lens} we present a 2D histogram showing GAME Filter's sensitivity expressed as the ratio \( N_{20\%}/N_\mathrm{all} \) across \( D_\mathrm{L} \) and \( M_\mathrm{L} \) parameter space. As expected, our algorithm demonstrates the highest sensitivity for nearby lenses with high masses, where the astrometric microlensing signal is strongest. For lenses with \( D_\mathrm{L} \lesssim 1\,\mathrm{kpc} \) and \( M_\mathrm{L} \gtrsim 1\,\mathrm{M_\odot} \), the sensitivity is \( 70\%-80\% \). For \( M_\mathrm{L} \sim 20\,\mathrm{M_\odot} \), the sensitivity drops to \(\approx 50\% \). This is caused by a large Einstein Radius \( \theta_\mathrm{E} \) and hence a very long event time-scale.
Consequently, longer observation times are required to capture the full event, limiting GAME Filter's ability to constrain the lens properties accurately within Gaia DR4 time frame. At distances of \( 4-6\,\mathrm{kpc} \), the sensitivity ranges from \( 10\%-20\% \) for lenses with \( M_\mathrm{L} \gtrsim 2\,\mathrm{M_\odot} \). For \( D_\mathrm{L} > 4\,\mathrm{kpc} \) and \( M_\mathrm{L} < 2\,\mathrm{M_\odot} \), the sensitivity drops below \( 10\% \), reflecting the significantly weaker astrometric microlensing signal.}

{The green and cyan circle symbols in Figure \ref{fig:lens} represent the first two microlensing events studied using combined photometry and astrometry: OB110462 \citep{Mroz_2022ApJ...937L..24M} and GDR3-001 \citep{Jablonska2022}. Both events lie within regions of high sensitivity, demonstrating the GAME Filter's ability to detect similar events.}

\subsection{Caveats}

{When generating astrometric events, we assumed a uniform distribution of their parameters, which does not accurately reflect their true distributions in the Milky Way (e.g. \citealt{Mao_1996}). However, the primary aim of our study is not to model the Galaxy accurately but to validate the performance and accuracy of GAME filter across a wide range of parameters. Using uniform distributions allowed us to test the filter under diverse conditions and identify specific regions in the parameter space where GAME filter may underperform. In such cases, deviations from a uniform recovery distribution can highlight areas where parameter estimation is less accurate or more difficult.}

The optimisation process in GAME Filter may occasionally converge to a local minimum instead of the global minimum, potentially leading to incorrect parameter estimates. To mitigate this, we used multiple initial guesses for parameters $u_0$, $\pi_\mathrm{EE}$, and $\pi_\mathrm{EN}$ for which we lack initial estimates (see Appendix \ref{app:game_filter}). Specifically, we used 10 different values for $u_0$ and 5 different values for both $\pi_\mathrm{EE}$ and $\pi_\mathrm{EN}$ within a defined interval. Although this approach improves parameter recovery accuracy, it does not cover the entire parameter space. To evaluate this, we further tested GAME Filter on a subset of \texttt{lens\_G14\_N281\_DR4} {mock dataset} with 5000 events, increasing the number of initial guesses for $u_0$, $\pi_\mathrm{EE}$ and $\pi_\mathrm{EN}$ to 13, 7, and 7, respectively. We also applied this approach to the parameters $\theta_\mathrm{E}$, $t_0$, and $t_\mathrm{E}$, where we considered 5 different values for individual parameters. We found no significant improvement compared to our original method. Due to this, we conclude that our original approach strikes a balance between computational efficiency, thorough parameter space exploration, and accurate parameter estimation.

Another limitation of GAME Filter is the assumption of negligible blending in the observed data. Blending occurs when light from the source combines with light from the lens (or/and a nearby unresolvable source), distorting the apparent motion of the light centroid and potentially leading to inaccurate microlensing parameter estimates. To accurately account for this effect, the blending parameter $f_\mathrm{bl}\in[0,1]$ is required. The measured Gaia parallax $\pi_\mathrm{G}$ in a simple case of source and luminous lens before or after the event (when the amplification is 1) can be expressed as
\begin{equation}
    \pi_\mathrm{G} = \varpi f_\mathrm{bl}+\pi_\mathrm{L}(1-f_\mathrm{bl}).
\end{equation}
and similarly for the proper motions. When $f_\mathrm{bl}$ is near 1, the observed signal mainly reflects the motion of the source, while values close to 0 reflect the motion of the lens. Intermediate values represent mixed cases where accurate estimation of the lens parallax requires additional source distance measurements, e.g., spectroscopic typing. In our study, we considered $f_\mathrm{bl}=1$ to determine the lower boundary for the Einstein radius $\theta_E$. This, in turn, establishes a minimum estimate for both the distance and mass of the lens. 
If $f_\mathrm{bl}$ were less than 1, indicating a significant contribution of light from the lens, the actual distance and mass would likely be higher than the values estimated under this assumption.

\section{Conclusions} \label{sec:conclusion}

The \textit{Gaia} mission provides an unprecedented opportunity to observe microlensing events solely as astrometric anomalies. Although photometric detection has its merits, this study focused on the identification of events only through the astrometric data. To leverage the astrometric precision of \textit{Gaia}, we developed GAME Filter, a software capable of identifying microlensing events and deriving the properties of the lensing objects that cause them. The main conclusions of our study are the following.
\begin{itemize}
    \item {GAME Filter can be used to characterise lenses with astrometry-only data for lens masses ranging from approximately 0.1 to 20$\,\mathrm{M_\odot}$ and distances up to $\approx 7\,$kpc. While the sensitivity decreases with increasing distance and decreasing lens mass, currently studied microlensing events fall within regions of high sensitivity.} 
    \item Values of MUWE and L2 optimality error after minimisation can be used to filter recovered events where GAME Filter reached the desired convergence.
    \item Recovery of true microlensing parameters is more challenging in specific regions of the parameter space, such as low $\theta_\mathrm{E}$ and high $|u_0|$, due to the weaker microlensing effect, resulting in higher scatter around true values and higher parameter estimation errors.
    \item Microlensing parameters are more difficult to recover for very short and very long duration events, since \textit{Gaia} might observe only a fraction of a microlensing event.
    \item There are no strong degeneracies for recovered events. Additionally, astrometry breaks the common degeneracy between the north and east components of the microlensing parallax present in photometry-only microlensing.
    \item As the magnitude increases, the observed signal becomes fainter, decreasing the number of recovered events and increasing the error of parameter estimation. We found that even for $G_0 = 19$, parameters can still be recovered for $\theta_\mathrm{E} \gtrsim 2\,${mas}.
    \item Observing regions with different numbers of \textit{Gaia} visits does not significantly affect the number and parameter estimation errors of recovered events, provided the number of visits is above average, i.e. $\gtrsim 90$.
    \item Even if the peak of the microlensing event is outside of the \textit{Gaia} observational run, microlensing parameters can still be recovered, although with higher errors.
    \item Binary systems represent a potential contaminant for GAME Filter's performance.  {We found that only $\approx 5$\% of binary events get falsely identified as weak microlensing events for $G_0=14$. However, we assumed uniform orbital parameter distributions, likely overestimating the confusion rate by underrepresenting short-period and overrepresenting long-period binaries. Nevertheless, we found that the false positives exhibit uniform distributions of parameters, suggesting that statistical noise rather than particular orbital configurations is the dominant cause of confusion.}
\end{itemize}

{We presented our results for \textit{Gaia} DR4, where GAME Filter demonstrates robust performance in detecting astrometric microlensing events. For DR5, we expect a similar overall performance. Additionally, we expect that the longer observing time will allow the detection of longer-duration events, particularly those with larger $\theta_\mathrm{E}$, which may have been missed in DR4 due to insufficient sampling. Such events are expected to be less frequent, as lenses with large masses are inherently rarer. However, these events would be of special astrophysical interest, as they are more likely to involve massive black holes}. {Additionally, we note that a quantitative estimate of the number of astrometric microlensing events detectable in DR4 requires detailed simulations with realistic Galactic population models. Such an estimate is beyond the scope of this paper. We plan to provide detailed predictions of event rates using realistic populations in a future study (Kaczmarek, et al., in prep.).}

The work presented here suggests that astrometric data can not only complement photometric detections but also enable microlensing detections and the determination of lens parameters independently, even in the absence of photometric signals. Therefore, this study paves the way for future astrometric investigations that fully leverage the capabilities of \textit{Gaia} data.

\section*{Data availability}

The software to generate mock \textit{Gaia} observations and GAME Filter are available at \href{https://github.com/tajjankovic/GAME-Filter/}{https://github.com/tajjankovic/GAME-Filter/}.

\begin{acknowledgements}
T. J., A. G., {\L}. W., T. P., M. K., and M. L. acknowledge the financial support from the European Space Agency PRODEX Experiment Arrangement GAME. T. J., A. G., T. P., M. B., and M. K. acknowledge the financial support from the Slovenian Research Agency (research core funding P1-0031, infrastructure program I0-0033, and project grants Nos. J1-8136, J1-2460, N1-0344). {\L}. W. acknowledges the funding from the European Union's research and innovation programme under grant agreements Nos. 101004719 (ORP) and 101131928 (ACME). We acknowledge ESA Gaia and the many researchers from Gaia DPAC for their valuable contributions and expertise at different phases of this project. Researcher T. J. conducts his research under the Marie Skłodowska-Curie Actions – COFUND project, which is co-funded by the European Union (Physics for Future – Grant Agreement No. 101081515). Z. K. is a Fellow of the International Max Planck Research School for Astronomy and Cosmic Physics at the University of Heidelberg (IMPRS-HD).

The following software was used in this work: 
    \textit{astromet} (\href{https://github.com/zpenoyre/astromet.py}{https://github.com/zpenoyre/astromet.py}), 
    \textit{jaxtromet} (\href{https://github.com/maja-jablonska/jaxtromet.py}{https://github.com/maja-jablonska/jaxtromet.py}), 
    \textit{jaxopt} \citep{jaxopt_implicit_diff}, 
    \textit{Matplotlib} \citep{Hunter:2007}.
    
\end{acknowledgements}

\bibliography{bibliography}

\begin{appendix}

\section{GAME Filter}\label{app:game_filter}

GAME Filter\footnote{\href{https://github.com/tajjankovic/GAME-Filter/}{https://github.com/tajjankovic/GAME-Filter/}} is a software tool developed to identify microlensing events in the \textit{Gaia} dataset and derive the properties of the lensing objects. GAME Filter processes $N$ events (the number of \textit{Gaia} data files, typically in \texttt{.parquet} format) in parallel. Each CPU or GPU core reads a file containing the data for an event. Specifically, the software reads $x_\mathrm{obs}$, $x_\mathrm{err}$, $\Delta x_\mathrm{obs}$, $t_\mathrm{obs}$, and $\varphi_\mathrm{obs}$. GAME Filter calculates $x_\mathrm{fit}$, the deviation along $\varphi_\mathrm{obs}$ at $t_\mathrm{obs}$, for specific single source and microlensing parameters. The software then minimises a scalar parameter MUWE (see Equation \ref{eq:muwe}), which indicates the goodness of the microlensing fit. The minimisation process utilises the L-BFGS-B algorithm to explore the parameter space and determine the optimal single source and microlensing parameters for individual events. It saves the best-fit results to disk and repeats these steps for every file.

In the following, we describe a method for obtaining initial guesses for individual parameters $\alpha_0^0$, $\delta_0^0$, $\mu_{\alpha^\star}^0$, $\mu_\delta^0$, $\varpi^0$,  $u_0^0$, $\theta_\mathrm{E}^0$, $t_0^0$, $t_\mathrm{E}^0$, $\pi_\mathrm{EE}^0$, and $\pi_\mathrm{EN}^0$, and the main steps of the minimisation process.

\subsection{Initial guess}

\begin{figure}
	\centering
	\begin{minipage}[b]{\linewidth}
		\includegraphics[width=\textwidth]{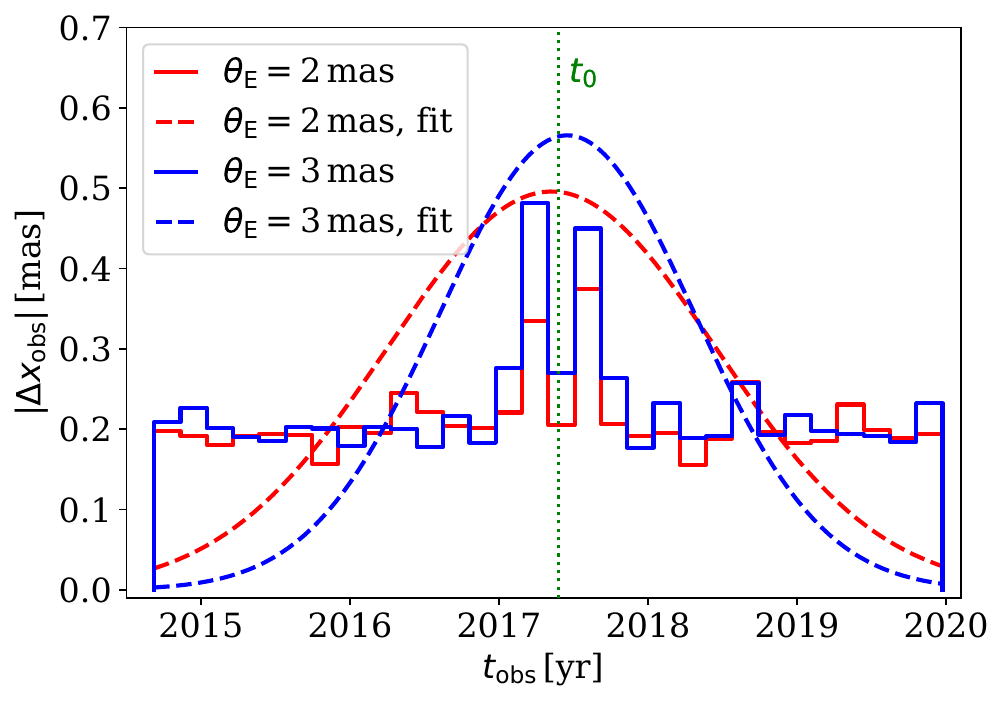}
	\end{minipage}
        \begin{minipage}[b]{\linewidth}
		\includegraphics[width=\textwidth]{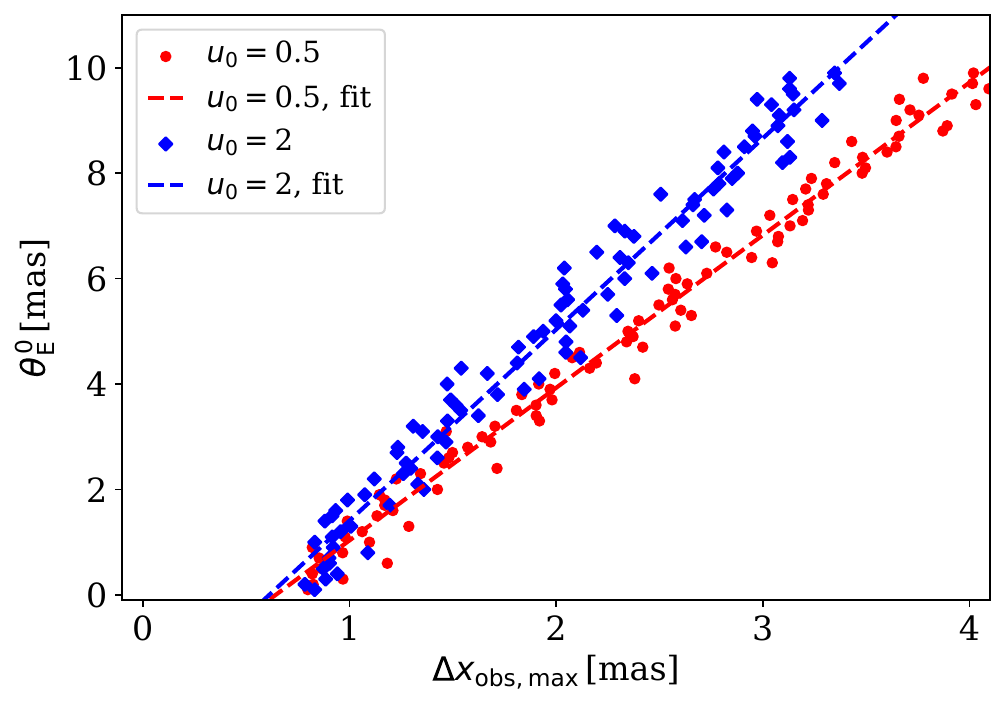}
	\end{minipage}
 
	\caption{{Upper panel}: Residuals calculated as the absolute value of the mean of individual bins (solid lines) for $\theta_\mathrm{E}=1\,$mas (red) and $\theta_\mathrm{E}=3\,$mas (blue). We used 30 equally spaced bins in $t_\mathrm{obs}$. Histograms are normalized to 1. We assume: $\alpha=6.49^\circ$, $\delta=-47.6^\circ$, $\varpi$=1, $\mu_\alpha^*=0$, $\mu_\delta=0$, $u_0=0.5$, $t_0=2017.3$ (denoted by a green dotted vertical line), $t_\mathrm{E}=30$ days, $\pi_\mathrm{EE}=\pi_\mathrm{EN}=0.001$. Residuals are fitted with a Gaussian (dashed lines).
{Lower panel}: $\theta_\mathrm{E}^0$ as a function of the maximum of residuals obtained from a Gaussian fit for $u_0=0.5$ and (red circle symbols) and $u_0=2$ (blue diamond symbols). The dashed lines denote the best linear fit of the scatter plots.}
	\label{fig:minimiz}
\end{figure}

The single source initial guess values for $\alpha_0^0$, $\delta_0^0$, $\mu_{\alpha^\star}^0$, $\mu_\delta^0$, $\varpi^0$ are obtained from the \textit{Gaia} files, corresponding to the values obtained from a single source fit. In addition, we used $u_0^0 = 0.01$, $\pi_\mathrm{EE}^0 = 0.01$, and $\pi_\mathrm{EN}^0 = 0.01$. We obtain the initial guesses $\theta_\mathrm{E}^0$, $t_0^0$, and $t_\mathrm{E}^0$ by fitting a Gaussian function to the histograms of residuals $\Delta x_\mathrm{obs}$, included in the \textit{Gaia} data files, using the \texttt{curve\_fit} function from the \texttt{scipy.optimize} module in the SciPy library. This is shown in Figure \ref{fig:minimiz} (top panel). We relate $\theta_\mathrm{E}^0$ to the maximum Gaussian $\Delta x_\mathrm{obs,max}$, $t_0^0$ to the time $t_\mathrm{obs}$ of the maximum of the Gaussian, and $t_\mathrm{E}^0$ to the standard deviation of the Gaussian.  We find that the relation between $\theta_\mathrm{E}$ and $\Delta x_\mathrm{obs,max}$ can be approximated as $\theta_\mathrm{E}^0 \approx 3  \Delta x_\mathrm{obs,max} - 2.25$. We show this correlation in Figure \ref{fig:minimiz} (lower panel) for different $u_0$. We see that the change in $u_0$ does not significantly affect the relation between $\theta_\mathrm{E}$ and $\Delta x_\mathrm{obs,max}$.

\newpage
\subsection{Minimisation steps}

We implemented several steps to improve the accuracy of GAME Filter. These criteria are based on the L2 optimality error $L_\mathrm{opt}$, initial guesses, and the boundaries imposed on individual parameters. $L_\mathrm{opt}$, often used in optimisation problems, determines the accuracy of the minimiser by defining the change during subsequent iterations in the parameter space, where values closer to 0 indicate a more accurate solution. We find that using minimisation methods
with the bounds and changing the initial guesses $u_0^0$, $\pi_\mathrm{EE}^0$, and $\pi_\mathrm{EN}^0$ significantly improves
the accuracy of GAME filter. Based on this, the main minimisation steps are as follows.

\begin{figure}
	\centering
        \begin{minipage}[b]{\linewidth}
\includegraphics[width=\textwidth]{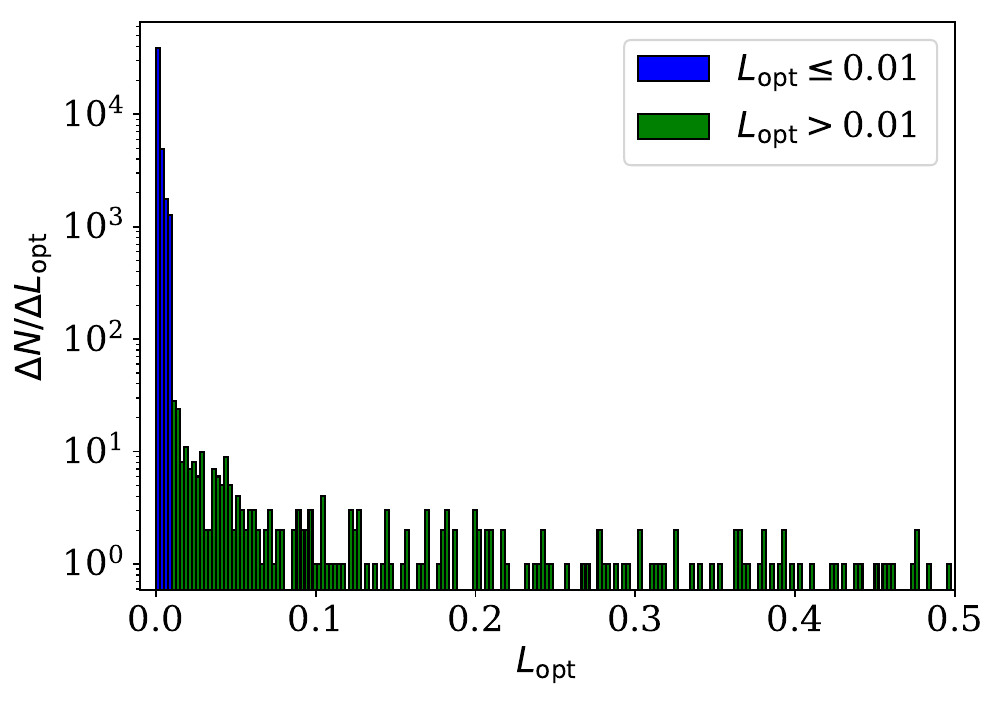}	\end{minipage}
	\caption{Distribution of $L_\mathrm{opt}$ for the \texttt{lens\_G14\_N281\_DR4} {mock dataset} (see Tables \ref{tab:micro_params} and \ref{tab:event_params}). Blue and green colours denote events with $L_\mathrm{opt}\leq0.01$ and $L_\mathrm{opt}>0.01$, respectively. The values in individual bins denote the number of events within the corresponding bin range.}
	\label{fig:Lopt}
\end{figure}

\begin{enumerate}
    \item Calculate $L_\mathrm{opt}$ after minimisation.
    \item If $L_\mathrm{opt}$ is higher than a threshold value $L_\mathrm{thresh}=0.01$, change $u_0^0$,
    $\pi_\mathrm{EE}^0$, and $\pi_\mathrm{EN}^0$. We determine $L_\mathrm{thresh}=0.01$ from $L_\mathrm{opt}$ histograms as the value where $90$\% of the data falls below this threshold, effectively removing the top $10$\% of values in the distribution tails. This is also shown in Figure \ref{fig:Lopt} (lower panel). 
    \begin{enumerate}
        \item For $u_0^0$, use 10 different values from an evenly spaced interval of $u_0^0 \in [-5,5]$.
        \item For $\pi_\mathrm{EE}^0$ and $\pi_\mathrm{EN}^0$, use 5 different values from an evenly spaced interval of $\pi_\mathrm{EE}^0, \pi_\mathrm{EN}^0 \in [-1,1]$.
    \end{enumerate}
    \item Stop the iteration if $L_\mathrm{opt} < L_\mathrm{thresh}$. Otherwise, keep the parameters with the lowest $L_\mathrm{opt}$.
\end{enumerate}

 \end{appendix}

\end{document}